\documentclass[twocolumn,a4paper,showpacs,superscriptaddress,pre,aps,floatfix]{revtex4-1}

 \usepackage{amsmath}
 \usepackage{amssymb}
 \usepackage{bbold}
 \usepackage{latexsym}
 \usepackage{amsfonts}
 \usepackage[caption=false]{subfig}
 \usepackage{epsfig}
 \usepackage{psfrag}
 \usepackage{color}
 \definecolor{darkblue}{rgb}{0,0,.5}
 \usepackage[linktocpage, colorlinks=true ,linkcolor=darkblue, citecolor=darkblue]{hyperref}
 \usepackage[all]{hypcap}
\usepackage{graphicx}

 \newcommand{\ket}[1]{\left|#1\right>}
 \newcommand{\bra}[1]{\left<#1\right|}
 \newcommand{\expval}[1]{\left< #1 \right>}
 
 \newcommand{\braket}[2]
 {\left<#1|#2\right>}
 \newcommand{\nn}{\nonumber\\}
 
 \newcommand{\f}[1]{\mbox{\boldmath$#1$}}

 \newcommand{\bea}{\begin{eqnarray}}
 \newcommand{\eea}{\end{eqnarray}}
 \newcommand{\be}{\begin{equation}}
 \newcommand{\ee}{\end{equation}}
 \newcommand{\ord}{{\cal O}}
 
 \newcommand{\traceB}[1]{{\rm Tr_B}\left\{ #1 \right\}}
 
 \newcommand{\abs}[1]{{\left| #1 \right|}}

 \newcommand{\ii}{{\rm i}}
\begin{document}

\title{Collective couplings: rectification and supertransmittance}

\author{Gernot Schaller}
\email{gernot.schaller@tu-berlin.de}
\affiliation{Institut f\"ur Theoretische Physik, Technische Universit\"at Berlin,
Hardenbergstra{\ss}e 36, D-10623 Berlin, Germany}
\author{Giulio Giuseppe Giusteri}
\affiliation{Mathematical Soft Matter Unit, Okinawa Institute of Science and Technology Graduate University, 1919-1 Tancha, Onna, Okinawa, 904-0495, Japan}
\affiliation{Dipartimento di Matematica e Fisica and ILAMP, Universit\`a Cattolica del Sacro Cuore, I-25121, Brescia, Italy}
\affiliation{Istituto Nazionale di Fisica Nucleare, sez.~Pavia, via Bassi 6, I-27100, Pavia, Italy}
\author{Giuseppe Luca Celardo}
\affiliation{Dipartimento di Matematica e Fisica and ILAMP, Universit\`a Cattolica del Sacro Cuore, I-25121, Brescia, Italy}
\affiliation{Istituto Nazionale di Fisica Nucleare, sez.~Pavia, via Bassi 6, I-27100, Pavia, Italy}
\affiliation{Instituto de F\'isica, Benem\'erita Universidad Aut\'onoma de Puebla, Apartado Postal J-48, Puebla 72570, Mexico}


\begin{abstract}
We investigate heat transport between two thermal reservoirs that are coupled via a large spin composed of $N$ identical two level systems.
One coupling implements the dissipative Dicke superradiance.
The other coupling is locally of the pure-dephasing type and 
requires to go beyond the standard weak-coupling limit by employing a 
Bogoliubov mapping in the corresponding reservoir.
After the mapping, the large spin is coupled to a collective mode with the original 
pure-dephasing interaction, but the collective mode is dissipatively coupled to the residual 
oscillators. 
Treating the large spin and the collective mode as the system, 
a standard master equation approach is now able to capture the energy transfer between the two reservoirs.
Assuming fast relaxation of the collective mode, we derive a coarse-grained rate equation for the large spin only and
discuss how the original Dicke superradiance is affected by the presence of the additional reservoir.
Our main finding is a cooperatively enhanced rectification effect due to
the interplay of supertransmittant heat currents (scaling quadratically with $N$)
and the asymmetric coupling to both reservoirs.
For large $N$, the system can thus significantly amplify
current asymmetries under bias reversal, functioning as a heat diode. 
{We also briefly discuss the case when the  couplings of the
  collective spin are locally dissipative, showing that the heat-diode
  effect is still present.}
\end{abstract}

\maketitle

\section{Motivation}

The study of radiative effects in two-level systems has a long history.
Here, the spin-boson model~\cite{leggett1987a} takes a very prominent role.
Originating from the interaction of a two-level atom with the
electromagnetic field~\cite{scully1997}, it is often used as a toy model in many other contexts.
Not surprisingly, it has become a canonical model to explore fundamental methods of open systems~\cite{wilhelm2004a,nesi2007a,wang2015a}
and effectively arises in a rather large number of physical 
systems and effects, including e.g.\ the dynamics of light-harvesting complexes~\cite{cheng2009a}, 
detectors~\cite{hegerfeldt2007a}, 
and the interaction of quantum dots with generalized environments~\cite{brandes2005a}. 

Ideally, one aims at a reduced description taking only the finite-dimensional spin
dynamics into account.
However, when the number of spins is increased, the curse of dimensionality -- the exponential 
growth of the system Hilbert space with its size -- usually inhibits investigations
of large spin-boson models.
When additional symmetries come into play -- e.g.\ when the spins have the same splitting and couple 
collectively to all other components -- simplified descriptions are applicable.
Collective effects may for example dramatically influence the dephasing behavior of the environment, leading to phenomena
such as super- and sub-decoherence~\cite{palma1996a,unruh1995a}.
Furthermore, they play a significant role in the modelling of light-harvesting 
complexes~\cite{celardo2012a,ferrari2014a,celardo2014a,celardo2014b}.
Perhaps one of the clearest manifestations of collective behaviour is Dicke superradiance~\cite{dicke1954a}.
Here, the collectivity of the coupling between $N$ two-level atoms and a low-temperature bosonic reservoir induces an 
unusually fast relaxation.
When the atoms couple independently, the time needed for relaxation does not depend on $N$.
For a collective coupling, however, the relaxation time scales as $1/N$
and the maximum radiation intensity scales as $N^2$~\cite{breuer2002}.
A setup in which the collective coupling approximation is well justified can be obtained 
by confining the two-level atoms in a region much smaller than
the wavelength of the electromagnetic field, but such collective couplings may also be engineered for instance using trapped ions~\cite{cormick2013a}
or opto-mechanical setups~\cite{mumford2015a}.

Transient superradiant phenomena have been investigated from many perspectives both theoretically~\cite{andreev1980a,gross1982a,brandes2005a}
and experimentally~\cite{flusberg1976a,mlynek2014a}.
Our present study is motivated by the fact that in certain regimes the transient superradiance can be
turned into a stationary supertransmittance -- a stationary current scaling with $N^2$ -- when two collective weakly-coupled 
reservoirs~\cite{vogl2011a} or a combination of
weak collective dissipation and driving~\cite{meiser2010a,meiser2010b} are considered.
Superradiance  has been studied also in
the context of single excitation transport~\cite{luca1,luca2,luca3,luca4}.

Naturally, when {reservoirs of the same nature} are coupled with the same operators to the system~\cite{vogl2011a,wang2015b}, 
general symmetry arguments suggest that, under reversal of the thermal bias, the heat current will simply revert its sign.
By contrast, when the reservoirs are coupled with different operators to the system, one may notice asymmetries in 
the heat currents under bias reversal.
Typically, these are not very pronounced~\cite{schaller2009b} and are often expected to
average out when multiple systems are used in parallel.
When the absolute value of the current is significant in one non-equilibrium configuration but is strongly suppressed under temperature exchange, 
one effectively implements a heat diode~\cite{segal2008b,ruokola2011a,ren2013a,li2015a}.
The ideal heat diode would display a very large heat conductivity in one bias configuration and a complete suppression in the
opposite.

We show that by implementing distinct collective couplings to source
and drain reservoirs {of different nature}, asymmetries in the heat conductance 
can be strongly amplified in the large-$N$ regime.

We present the model in Sec.~\ref{SEC:model} below, where, in particular, we also 
discuss the mapping to a collective reservoir mode in Sec.~\ref{SEC:bogoliubov_mapping}
and possible implementation scenarios in Sec.~\ref{SEC:implementations}.
For consistency, we also briefly recall the main features of superradiant decay in Sec.~\ref{SEC:superradiance}.
Then, we derive the quantum-optical master equation and discuss its
thermodynamic properties in Sec.~\ref{SEC:mastereq}.
We obtain a coarse-grained description for the large spin dynamics and investigate the 
modification of Dicke superradiance in Sec.~\ref{SEC:decay} and the resulting heat currents
between the reservoirs in Sec.~\ref{SEC:current}.
{In this article, by reservoirs of different nature, we mean that
  one is a standard heat bath made of a collection of harmonic
  oscillators, while the other reservoir is structured: it can be described by a single
  mode strongly coupled to the spin ensemble and at the same time coupled
  to an independent heat-bath. The coupling between the single mode
  and the spin ensemble can be locally of pure dephasing or
  dissipative type. We will mainly consider the pure-dephasing case since it
  is more tractable analytically. Nevertheless, also the case of
  dissipative coupling is discussed in Sec.~\ref{SEC:rectification},
  showing that the main result of our paper, namely the cooperative
  rectification effect, still applies.} 
A summary of our results is given in  in Sec.~\ref{SEC:summary}.

\section{Model}~\label{SEC:model}

\subsection{Hamiltonian}\label{SEC:hamiltonian}

Our model is described by the total Hamiltonian $H=H_S+H_I+H_B$, which we decompose in a standard way 
into a system, an interaction, and a reservoir (bath) contribution, respectively.
The system Hamiltonian is described by a large spin that is collectively (i.e., also via large-spin interactions)
coupled to a harmonic oscillator
\bea\label{EQ:ham_sys}
H_S = \frac{\omega_0}{2} J_z + \lambda J_z \otimes (a+a^\dagger) + \Omega a^\dagger a\,,
\eea
where $\omega_0$ denotes the level splitting of the large-spin, $\Omega$ the splitting of the harmonic oscillator, 
and $\lambda$ represents the coupling between them.
The large spin operator can be implemented by considering $N$ identical two-level systems described by 
Pauli matrices, such that $J_{x/y/z} = \sum\limits_{i=1}^N \sigma_{x/y/z}^i$.
Both the large spin and the harmonic oscillator are coupled to separate reservoirs $H_I = H_I^t+H_I^\ell$, where
\bea\label{EQ:ham_int}
H_I^t &=& J_x \sum_k \left(h_{kt} b_{kt} + h_{kt}^* b_{kt}^\dagger\right)\,,\nn
H_I^\ell &=& a \sum_k h_{k\ell} b_{k\ell} + a^\dagger \sum_k h_{k\ell}^* b_{k\ell}^\dagger\,,
\eea
where the $h_{k\nu}$ denote the bare emission/absorbtion amplitude for mode $k$ in reservoir $\nu$.
As the $b_{kt}$-modes couple in a direction that is transverse to the large spin Hamiltonian, we will in the following
call the associated reservoir the {\em transverse reservoir}, and the other reservoir the {\em longitudinal reservoir}.
The reservoirs are modeled as non-interacting oscillators
\bea\label{EQ:ham_res}
H_B^t &=& \sum_k \omega_{kt} b_{kt}^\dagger b_{kt}\,,\nn
H_B^\ell &=& \sum_k \omega_{k\ell} b_{k\ell}^\dagger b_{k\ell}\,,
\eea
which will be assumed to remain at thermal equilibrium states in the subsequent analysis.

Our system is thus completely defined in terms of the Hamiltonians in Eqns.~(\ref{EQ:ham_sys}), (\ref{EQ:ham_int}), and~(\ref{EQ:ham_res}).
The reader mainly interested in our results for this system may directly proceed to Sec.~\ref{SEC:mastereq}.
However, in Sec.~\ref{SEC:bogoliubov_mapping} below, we demonstrate that our model arises when the 
large spin is directly coupled to a longitudinal reservoir by treating a collective reservoir degree of 
freedom as part of the system.
Furthermore, we provide some hints about a possible physical implementation in Sec.~\ref{SEC:implementations}
and also review the Dicke model dynamics arising in the limit $\lambda\to 0$ in Sec.~\ref{SEC:superradiance}.

We would like to stress that the main findings of our
paper are recovered also when the coupling to the bosonic mode is implemented
by a locally dissipative interaction instead of a locally purely dephasing
interaction, i.e.\ substituting $\lambda J_z \rightarrow \lambda J_x$ in 
Eq.~(\ref{EQ:ham_sys}), as discussed in Sec.~\ref{SEC:rectification}. 
Moreover, we remark that, even when considering only the large spin as the system, 
the single bosonic mode together with its coupling~(\ref{EQ:ham_int}) to its heat bath~(\ref{EQ:ham_res}) 
represents a structured reservoir which is non-Markovian by construction. 
Indeed, the single bosonic mode is
dynamically evolving in our model and will adapt to the state of its connected heat bath.

\subsection{Explicit Bogoliubov mapping}\label{SEC:bogoliubov_mapping}

Thinking of the large-spin operators as implemented by identical two-level atoms that couple collectively to
photons and phonons, respectively, it seems more in line with traditional approaches to consider them to be
directly coupled to two reservoirs.
That is, the total Hamiltonian is now given by $H=\tilde{H}_S + \tilde{H}_I + \tilde{H}_B$.
We note that in this section, we mark all contributions that are different from the presentation in the previous section with a tilde.
Specifically, now the system is only described by the large spin
\bea\label{EQ:ham_sys1}
\tilde{H}_S = \frac{\omega_0}{2} J_z\,,
\eea
which is directly coupled to two reservoirs $\tilde{H}_I = H_I^t + \tilde{H}_I^\ell$
\bea\label{EQ:ham_int1}
H_I^t &=& J_x \otimes \sum_k \left(h_{kt} b_{kt} + h_{kt}^* b_{kt}^\dagger\right)\,,\nn
\tilde{H}_I^\ell &=& J_z \otimes \sum_k \left(\tilde{h}_{k\ell} \tilde{b}_{k\ell} + \tilde{h}_{k\ell}^* \tilde{b}_{k\ell}^\dagger\right)\,.
\eea
The two reservoirs $\tilde{H}_B = H_B^t + \tilde{H}_B^\ell$ are described by otherwise non-interacting oscillators
\bea\label{EQ:ham_res1}
H_B^t &=& \sum_k \omega_{kt} b_{kt}^\dagger b_{kt}\,,\nn
\tilde{H}_B^\ell &=& \sum_k \tilde{\omega}_{k\ell} \tilde{b}_{k\ell}^\dagger \tilde{b}_{k\ell}\,.
\eea
%
Starting from such a setup, we note that, since the interaction Hamiltonian of the longitudinal reservoir commutes with the system 
Hamiltonian $[\tilde{H}_S, \tilde{H}_I^{\ell}]={\bf 0}$, the longitudinal reservoir and the large spin system will not directly exchange energy.
Naively, one might then be tempted to believe that the model cannot support stationary heat currents from one reservoir to the other.
Nevertheless, a direct exchange of energy between the reservoirs is possible as the individual
interaction Hamiltonians do not commute $[\tilde{H}_I^\ell, H_I^t]\neq {\bf 0}$.
This effect is, however, of higher order
and thus beyond the reach of a naive master equation approach.
Generally, an interaction that appears to be locally of the pure-dephasing type as the second of Eqns.~(\ref{EQ:ham_int1}) 
need no longer preserve the energy of the system when further interactions are added.

We do now consider a Bogoliubov transformation of the longitudinal reservoir modes
\bea\label{EQ:bogoliubov}
\tilde{b}_{k\ell} = u_{k1} a + \sum_{k'>1} u_{kk'} b_{k'\ell}
\eea
to new bosonic operators $a$ and $b_{k\ell}$ with transformation coefficients $u_{kk'}$ yet to 
be determined.
We have written these new operators without a tilde symbol as we will demonstrate in the following that they
correspond to the operators used in Sec.~\ref{SEC:hamiltonian}.
We note that provided we start off with $\tilde{K}$ modes in the longitudinal reservoir, we will thus separate one mode $a$ from the corresponding
reservoir where only $K=\tilde{K}-1$ modes -- often called residual oscillators -- remain.
Requiring the new operators to fulfil the bosonic commutation relations yields the equation
\bea\label{EQ:bogoliubov1}
\delta_{kq} = u_{k1} u_{q1}^* + \sum_{k'>1} u_{kk'} u_{qk'}^*\,.
\eea
Normally, Bogoliubov transformations are applied to diagonalize a quadratic Hamiltonian. 
In contrast, here we only intend to change the form of the coupling.
Specifically, we require that the large spin should only couple to the collective degree of freedom described by mode $a$ and 
that the part of the Hamiltonian describing only the residual oscillators is diagonal, i.e., 
\bea\label{EQ:ham_desired}
\tilde{H}_I^\ell + \tilde{H}_B^\ell &\stackrel{!}{=}& J_z (\lambda a + \lambda^* a^\dagger) 
+ a \sum_{k>1} h_{k\ell}^* b_{k\ell}^\dagger + a^\dagger \sum_{k>1} h_{k\ell} b_{k\ell}\nn
&&+ \sum_{k>1} \omega_{k\ell} b_{k\ell}^\dagger b_{k\ell} + \Omega a^\dagger a\,.
\eea
Inserting the transformation~(\ref{EQ:bogoliubov}) and comparing with the desired form above, we find that it gives rise to two further constraints
\bea\label{EQ:bogoliubov2}
0 &=& \sum_k \tilde{h}_{k\ell} u_{kk'}\; : \; \forall \; k'>1\,,\nn
\omega_{k'\ell} \delta_{k'q'} &=& \sum_k \tilde{\omega}_{k\ell} u_{kk'}^* u_{kq'} \; : \; \forall \; k',q'>1\,.
\eea
Once we fulfil Eqns.~(\ref{EQ:bogoliubov1}) and~(\ref{EQ:bogoliubov2}), the coupling between the longitudinal reservoir and the large spin will 
be mediated by the collective coordinate $a$.
These conditions can be written as the problem of diagonalizing a hermitian matrix.
Defining the vectors
\bea
\ket{W_1} &=& \left(\begin{array}{c}
u_{11}\\
\vdots\\
u_{N1}
\end{array}\right)\,,\qquad
\ket{W_{k>1}} = \left(\begin{array}{c}
u_{1k}\\
\vdots\\
u_{Nk}
\end{array}\right)\,,\nn
\ket{H} &=& \frac{1}{\sqrt{\sum_k \abs{\tilde{h}_{k\ell}}^2}} 
\left(\begin{array}{c}
\tilde{h}_{1\ell}^*\\
\vdots\\
\tilde{h}_{N\ell}^*
\end{array}\right)\,,
\eea
we see that Eq.~(\ref{EQ:bogoliubov1}) can be satisfied by orthonormality of the vectors $\ket{W_k}$.
Furthermore, the first of Eq.~(\ref{EQ:bogoliubov2}) can be written as orthogonality between $\ket{H}$ and
all but the first previously defined vectors $\braket{H}{W_{k'>1}}=0$.
Finally, by defining the hermitian matrix
\bea
B = \left(\f{1}-\ket{H}\bra{H}\right) 
\left(\begin{array}{ccc}
\tilde{\omega}_{1\ell}\\
&\ddots\\
& & \tilde{\omega}_{\tilde{K}\ell}
\end{array}\right)
\left(\f{1}-\ket{H}\bra{H}\right)\,,\nn
\eea
we see that we can simultaneously satisfy Eqns.~(\ref{EQ:bogoliubov1}) and~(\ref{EQ:bogoliubov2}) when choosing the $\ket{W_k}$ as eigenvectors
of the matrix $B$.
By construction, $\ket{W_1}=\ket{H}$ is already one eigenvector with eigenvalue $0$.
The remaining eigenvectors are then for $k>1$ given by $B \ket{W_{k}} = \omega_{k} \ket{W_{k}}$.
As the matrix $B$ is hermitian, such a mapping can always be found and is exact.
The Hamiltonian then assumes the form of Eq.~(\ref{EQ:ham_desired})
with the explicit relations
\bea
\lambda &=& \sum_k \tilde{h}_{k\ell} u_{k1} = \frac{\sum_k \abs{\tilde{h}_{k\ell}}^2}{\sqrt{\sum_k \abs{\tilde{h}_{k\ell}}^2}} = \lambda^*\,,\nn
\Omega &=& \sum_k \tilde{\omega}_{k\ell} u_{k1}^* u_{k1} = \frac{\sum_k \tilde{\omega}_{k\ell} \abs{\tilde{h}_{k\ell}}^2}{\sum_k \abs{\tilde{h}_{k\ell}}^2} > 0\,,\nn
h_{k\ell} &=& \sum_{k'} \tilde{\omega}_{k'\ell	} u_{k'1}^* u_{k'k}\;:\;\forall\;k>1\,.
\eea
We therefore note that $\tilde{H}_S + \tilde{H}_I^\ell + \tilde{H}_B^\ell = H_S + H_I^\ell + H_B^\ell$, which exactly maps the total Hamiltonian of
Eqns.~(\ref{EQ:ham_sys1}), (\ref{EQ:ham_int1}), and~(\ref{EQ:ham_res1}) into the total Hamiltonian of  Eqns.~(\ref{EQ:ham_sys}), (\ref{EQ:ham_int}), and (\ref{EQ:ham_res}).
We also note that in the strong-coupling limit $\tilde{h}_{k\ell}\to\infty$, the renormalized frequencies $\Omega$ and $\omega_{k\ell}$ as
well as the renormalized couplings $h_{k\ell}$ remain finite.

\begin{figure}[t]
\includegraphics[width=0.38\textwidth,clip=true]{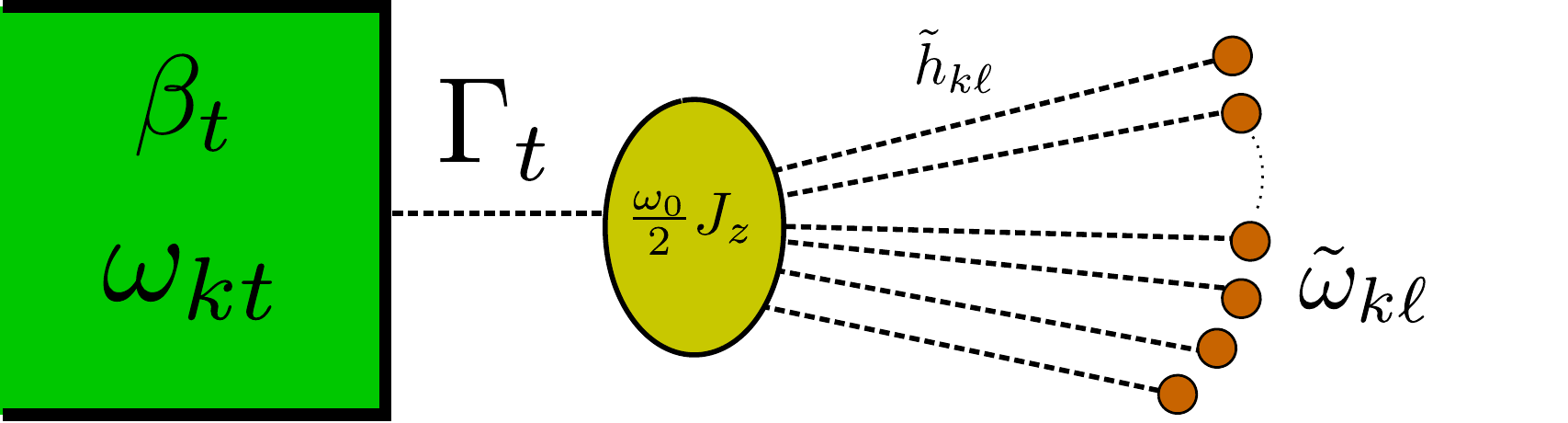}\\
\includegraphics[width=0.38\textwidth,clip=true]{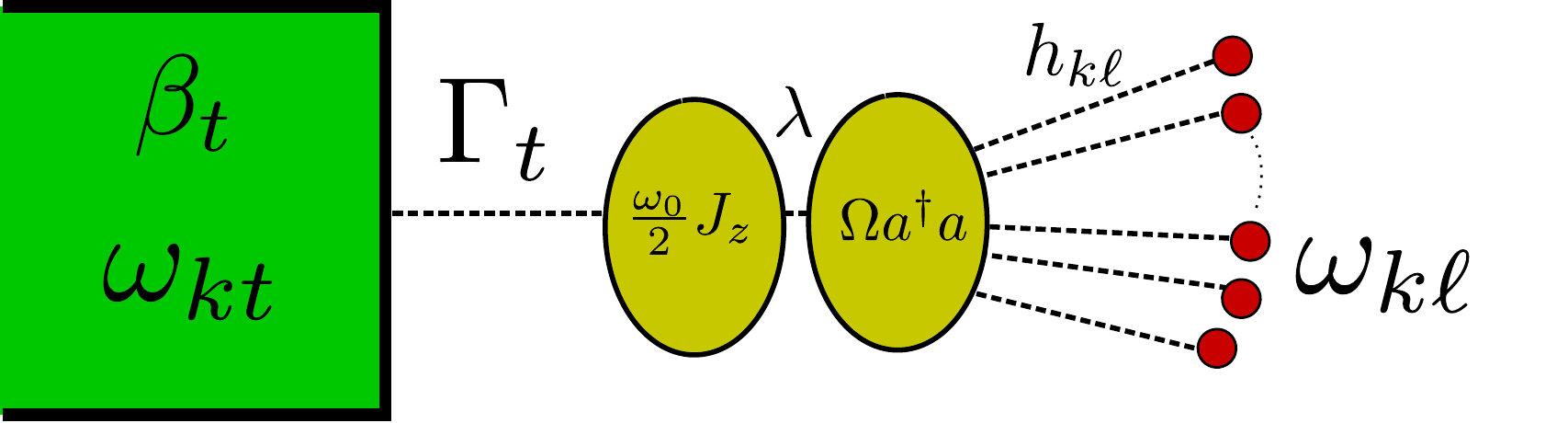}\\
\includegraphics[width=0.38\textwidth,clip=true]{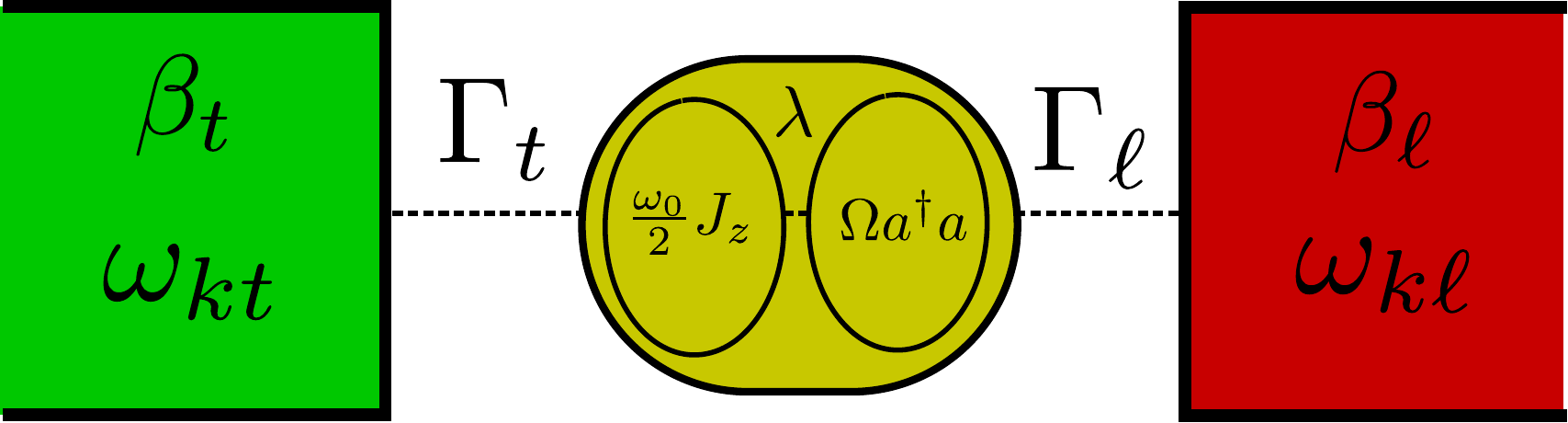}
\caption{\label{FIG:sketch_model}(Color Online)
Sketch of the described mapping procedure.
{\bf Top}: In the original system  -- compare Eqns.~(\ref{EQ:ham_sys1}), (\ref{EQ:ham_int1}), and~(\ref{EQ:ham_res1}) -- the large spin is directly
coupled to a transversal reservoir (via $J_x$, left and green) and a longitudinal reservoir (via $J_z$, right and orange).
{\bf Middle}: The Bogoliubov transformation applied on the longitudinal reservoir allows to separate a collective degree of freedom (described
by the oscillator mode $a$, center-right and yellow) from the longitudinal reservoir, leading to renormalized coupling strengths and energies, compare 
Eqns.~(\ref{EQ:ham_sys}), (\ref{EQ:ham_int}), and~(\ref{EQ:ham_res}).
{\bf Bottom}:
After the transformation, we apply a new decomposition into system and reservoirs (solid, colored), which allows one to
explore the limit of strong $\lambda$, while both reservoirs are considered in the continuum limit with 
the respective couplings $\Gamma_t$ and $\Gamma_\ell$ to the transversal and the residual longitudinal 
reservoirs -- cf. Eq.~(\ref{EQ:tunnelrates}) -- being treated perturbatively.
}
\end{figure}

In what follows, we assume that the above mapping to a collective mode has been performed and we will therefore consider
Eqns.~(\ref{EQ:ham_sys}),~(\ref{EQ:ham_int}), and~(\ref{EQ:ham_res}) as the starting point of our considerations.
Fig.~\ref{FIG:sketch_model} illustrates the effect of the applied mapping.
After transformation, we note that the coupling to both reservoirs is dissipative already to lowest order in the tunneling rates
\bea\label{EQ:tunnelrates}
\Gamma_t(\omega) &=& 2\pi \sum_k \abs{h_{kt}}^2 \delta(\omega-\omega_{kt}) \to \Gamma_t\,,\nn
\Gamma_\ell(\omega) &=& 2\pi \sum_k \abs{h_{k\ell}}^2 \delta(\omega-\omega_{k\ell}) \to \Gamma_\ell\,,
\eea
i.e., it does not commute with the system Hamiltonian~(\ref{EQ:ham_sys}).
By contrast, before the transformation the energy exchange between the reservoirs was a higher-order effect.
Similar mappings are frequently used in the literature to treat strong-coupling limits.
When they only involve position operators, the collective mode is then called reaction-coordinate~\cite{strasberg2016a,iles_smith2014a,martinazzo2011a}, 
but also mappings to different lattice topologies exist~\cite{huh2014a}.

The new reservoirs will be assumed to remain at their local thermal equilibrium states throughout and we will investigate the dynamics
of the system subject to these two environments. 
Now, a master equation treatment may cover the effects of a strong coupling $\lambda$ and may therefore
also predict stationary currents between the two reservoirs.
By contrast, for a vanishing coupling $\lambda=0$ the model is split into two independent
components, where the large spin coupled to the transversal reservoir represents the usual Dicke model with its
well-known superradiant behaviour.
We will use this Dicke regime as a benchmark test for our model, compare Sec.~\ref{SEC:superradiance}.

\subsection{Implementations}\label{SEC:implementations}

In the previous section, we have seen that a collective reservoir mode may always be introduced at the price of obtaining
renormalized energies and couplings.
It remains to motivate why the coupling between the two-level systems and the reservoirs should be identical, such that
large-spin operators arise, and the system can be treated in the simple angular momentum basis.

One way of achieving collective couplings as in Eqns.~(\ref{EQ:ham_sys}),~(\ref{EQ:ham_int}), and~(\ref{EQ:ham_res}) 
would be their effective engineering in a quantum simulator~\cite{dimer2007a,nagy2010a,mumford2015a,rotondo2015a}.
However, collective couplings also naturally arise when the distance between the two-level systems is much smaller than the wavelength
of the reservoir modes with which they interact.
A most natural example for this situation is a Bose-Einstein condensate of two-level systems, where all atoms occupy the 
same quantum state and therefore see no difference in their interaction with additional systems.
Indeed, it has recently been experimentally possible to implement the collective $J_x$-operator arising in the Dicke model
by placing a Bose-Einstein condensate in a cavity~\cite{baumann2010a}.

Another possible scenario could be when the two-level systems are represented by identical ions in a trap inside a cavity.
Here, the two internal states would be represented by electronic degrees of freedom, and the photons in the surrounding 
cavity would assume the role of the transversal reservoir.
Collectivity of the transversal coupling could then be achieved when the physical distance between the ions is much smaller 
than the diameter of the cavity.  
The other bosonic modes would correspond to the ions motional degrees of freedom in the trap potential as is usually done
in current experiments~\cite{jurcevic2014a}.
Specifically, the collective vibrational mode could represent the single longitudinal mode $a$ in Eq.~(\ref{EQ:ham_sys}).
Then, the residual longitudinal reservoir would consist of the other vibrational modes of the ions (relative motion). 

In reality, we note that phonons can be expected to couple not only along the longitudinal direction, i.e., in 
Eqns.~(\ref{EQ:ham_sys}) and~(\ref{EQ:ham_int1}) one would rather expect couplings of the form 
$J_z \otimes \left[\ldots\right] \to (\f{n} \cdot \f{J}) \otimes \left[\ldots\right]$ with normal vector $\f{n}$.
We expect our model to be valid when the dephasing resulting from the longitudinal component $n_z$ of their 
coupling dominates the dynamics~\cite{palma1996a}. 
Moreover, in Sec.~\ref{SEC:rectification} we also consider the case of a coupling in the
$x$ direction ($\f{n}=\f{e}_x$).
Furthermore, even when these conditions for collectivity are not strictly fulfilled, collective couplings may nevertheless arise in an effective
picture, when interactions between the two-level systems are much stronger than the perturbations induced by the reservoirs.
Such an effective picture could arise similar to Sec.~\ref{SEC:bogoliubov_mapping}, but now involving mappings between spin operators only, 
which may not necessarily obey the same simple algebra as the large spins.

\subsection{Superradiance for $\lambda=0$}\label{SEC:superradiance}

The original Dicke Hamiltonian 
\bea
H_D= \tilde{H}_S + H_I^t + H_B^t
\eea
is recovered as an isolated part of the total system when $\lambda \to 0$. 
For $H_D$ it is known that, when all spins are prepared in the most excited state and the temperature of the 
reservoir is small $\beta_t\omega_0 \gg 1$, 
the two-level systems will decay collectively, resulting in a sharply localized flash of radiation with a maximum intensity
scaling as $N^2$ and a width scaling as $1/N$~\cite{dicke1954a,breuer2002}.

At finite temperatures, the master equation for the spin system is given by~\cite{agarwal1973a,vogl2011a}
\bea\label{EQ:masterdicke}
\dot{\rho} &=& -\ii \bigg[\frac{\omega_0}{2} J_z, \rho\bigg] + \Gamma_t n_t \left[J_+ \rho J_- - \frac{1}{2} \left\{J_- J_+, \rho\right\}\right]\nn
&& + \Gamma_t (1+n_t) \left[J_- \rho J_+ - \frac{1}{2} \left\{J_+ J_-, \rho\right\}\right]\,,
\eea
where $\Gamma_t = 2 \pi \sum_k \abs{h_{kt}}^2 \delta(\omega_0-\omega_{kt})$ denotes the bare absorbtion and emission rate and
$n_t = (e^{\beta_t\omega_0}-1)^{-1}$ the Bose-Einstein distribution of the transversal boson reservoir with inverse temperature $\beta_t$, 
evaluated at the system transition frequency $\omega_0$.
Furthermore, we have used the collective ladder operators $J_\pm = (J_x \pm \ii J_y)/2$.
We can clearly identify the terms accounting for the closed spin evolution (commutator) and for the emission [$\propto\Gamma_t(1+n_t)$] or absorbtion [$\propto\Gamma_t n_t$] of
bosons by the large spin.
The ratio of these rates yields the simple Boltzmann factor since $n_t/(1+n_t) = e^{-\beta_t \omega_0}$, such that this master equation obeys the usual 
detailed balance relation, which leads to thermalization at finite reservoir temperatures.
In particular, in the standard Dicke limit ($n_t\to 0$) it predicts the
collective decay from the most excited state $m=+N/2$ into the ground state $m=-N/2$.

To diagonalize the spin part of the Hamiltonian we recall the angular momentum eigenstates (the length of the angular momentum is fixed $j=N/2$ and will be omitted)
\bea\label{EQ:angular_momentum_eigenstates}
J_z \ket{m} = 2 m \ket{m}\,,
\eea
where $m\in\{-\frac{N}{2},-\frac{N}{2}+1,\ldots,+\frac{N}{2}-1,+\frac{N}{2}\}$.
On these eigenstates, the $J_\pm$ operators act as
\bea\label{EQ:ladder_operators}
J_\pm \ket{m} = \sqrt{\frac{N}{2}\left(\frac{N}{2}+1\right)-m(m\pm 1)} \ket{m\pm 1}\,.
\eea

We can use these relations to represent the master equation~(\ref{EQ:masterdicke}) as a simple rate equation in the spin energy eigenbasis $\ket{m}$
(with $P_m = \bra{m}\rho\ket{m}$)
\bea\label{EQ:rate_equation}
\dot P_m &=& - \left[\Gamma_t n_t M_m^+ + \Gamma_t(1+n_t) M_m^- \right] P_m\nn
&& + \Gamma_t n_t M_m^- P_{m-1} + \Gamma_t (1+n_t) M_m^+ P_{m+1}\,,\nn
M_m^\pm &=& \frac{N}{2}\left(\frac{N}{2}+1\right)-m(m\pm1)\,.
\eea
The coherences between different energy eigenstates will (if initially present at all) evolve independently and simply decay 
(as is often the case for the standard quantum-optical master equation).
However, we note that the matrix elements $M_m^\pm$ entering the transition rates scale quadratically with the number of two-level systems $N$, 
which is the formal reason for the superradiant decay into the vacuum state.

\section{Master Equation}\label{SEC:mastereq}

In this section, we will now consider the case of finite coupling ($\lambda \neq 0$) between large spin and longitudinal boson.

\subsection{Transition Rates}

We treat the large spin and the longitudinal boson mode as the system, defined by the parameters
$\omega_0$, $\lambda$, and $\Omega$ in Eq.~(\ref{EQ:ham_sys}).
Provided that the spectrum of the system is non-degenerate (at least between admitted transitions), the
quantum-optical master equation~\cite{breuer2002,schaller2014} becomes a rate equation connecting only the populations
in the system energy eigenbasis $H_S\ket{a} = E_a \ket{a}$.
For an interaction Hamiltonian of the form $H_I = \sum_\alpha A_\alpha \otimes B_\alpha$ with system operators 
$A_\alpha$ and reservoir operators $B_\alpha$, respectively, the 
rates from eigenstate $b$ to eigenstate $a$ are formally given by~\cite{schaller2014}
\bea\label{EQ:rate_general}
\gamma_{ab,ab} = \sum_{\alpha\beta} \gamma_{\alpha\beta}(E_b-E_a) \bra{a} A_\beta \ket{b} \bra{a} A_\alpha^\dagger \ket{b}^*\,,
\eea
where 
\bea\label{EQ:corrfunc_general}
\gamma_{\alpha\beta}(\omega) = \int e^{+\ii\omega\tau} \traceB{\f{B_\alpha}(\tau) B_\beta \rho_B} d\tau
\eea
are Fourier transforms of the reservoir correlation functions (bold symbols denote the interaction picture throughout).
Since the reservoir state
\bea
\rho_B = \frac{e^{-\beta_\ell(H_\ell - \mu_\ell N_\ell)}}{Z_\ell} \otimes \frac{e^{-\beta_t(H_t - \mu_t N_t)}}{Z_t}
\eea
is a tensor product of the thermal individual reservoir states, the temperatures enter the reservoir correlation 
functions $\gamma_{\alpha\beta}(\omega)$, 
whereas the collective coupling properties enter the matrix elements of the system coupling operators $A_\alpha$.

Specifically, we can identify in our interaction Hamiltonian~(\ref{EQ:ham_int}) the coupling operators 
$A_1 = J_x$, $A_2 = a$, $A_3 = a^\dagger$, $B_1 = \sum_k \left(h_{kt} b_{kt} + h_{kt}^* b_{kt}^\dagger\right)$,
$B_2 = \sum_k h_{k\ell} b_{k\ell}$, and $B_3 = \sum_k h_{k\ell}^* b_{k\ell}^\dagger$.
Consequently, the Fourier transforms of the non-vanishing correlation functions become
\bea\label{EQ:ft_corrfunc}
\gamma_{11}(\omega) &=& \Theta(+\omega) \Gamma_{t}(+\omega) [1+n_t(+\omega)]\nn
&&+ \Theta(-\omega) \Gamma_{t}(-\omega) n_t(-\omega)\,,\nn
\gamma_{23}(\omega) &=& \Theta(+\omega) \Gamma_{\ell}(+\omega) [1+n_\ell(+\omega)]\nn
\gamma_{32}(\omega) &=& \Theta(-\omega) \Gamma_\ell(-\omega) n_\ell(-\omega)\,,
\eea
where $\Gamma_{t/\ell}(\omega) = 2\pi \sum_k \abs{h_{kt/\ell}}^2 \delta(\omega-\omega_{kt/\ell})$ denotes the spectral coupling 
density of transversal and longitudinal reservoirs, 
$\Theta(\omega)$ the Heaviside step function, and
\bea
n_{\nu}(\omega) = \frac{1}{e^{\beta_\nu(\omega-\mu_\nu)}-1}
\eea
the Bose distribution of reservoir $\nu\in\{t,\ell\}$ with inverse temperature $\beta_\nu$ and chemical
potential $\mu_\nu \leq 0$.
In this paper, we will consider the case of vanishing chemical potentials $\mu_\nu=0$, 
but results expressed in terms of $n_\nu(\omega)$ will also hold for finite chemical potentials 
(used to effectively model interactions between bosons).

\subsection{Energy Eigenbasis}

To obtain the eigenbasis of (\ref{EQ:ham_sys}), we find it useful to employ the polaron transformation~\cite{mahan2000}
\bea\label{EQ:polaron_transformation}
U = e^{J_z B}\,,\qquad B = \frac{\lambda^*}{\Omega} a^\dagger - \frac{\lambda}{\Omega} a\,,
\eea
which -- since $B^\dagger = -B$ is anti-Hermitian -- acts unitarily on the operators.
It is straightforward to show the following relations
\bea\label{EQ:polaron_action}
U a U^\dagger &=& a - \frac{\lambda^*}{\Omega} J_z\,,\qquad U a^\dagger U^\dagger = a^\dagger - \frac{\lambda}{\Omega} J_z\,,\nn
U J_+ U^\dagger &=& J_+ e^{+2 B}\,,\qquad U J_- U^\dagger = J_- e^{-2 B}\,.
\eea
Consequently, the polaron transformation can be used to effectively decouple 
spin and polaron mode
\bea
\tilde H_S = U H_S U^\dagger = \frac{\omega_0}{2} J_z - \frac{\abs{\lambda}^2}{\Omega} J_z^2 + \Omega a^\dagger a\,.
\eea
The eigenstates $\widetilde{\ket{n,m}} = \widetilde{\ket{n}} \otimes \widetilde{\ket{m}}$ 
of $\tilde H_S$ are tensor products of the conventional angular momentum eigenstates~(\ref{EQ:angular_momentum_eigenstates}) and the Fock states
$\widetilde{\ket{n}}$, where we note that the conventional relations for creation and annihilation operators hold in the polaron-basis, 
e.g.\ $a^\dagger \widetilde{\ket{n}} = \sqrt{n+1} \widetilde{\ket{n+1}}$.
Consequently, the eigenstates $\ket{n,m}$ of $H_S$ can also be labeled by the spin quantum number $m\in\{-N/2,\ldots,+N/2\}$ and the occupation number
of the longitudinal boson mode $n\in\{0,1,2,\ldots\}$
and have energies
\bea
E_{nm} = \Omega n + \omega_0 m - 4 \frac{\abs{\lambda}^2}{\Omega} m^2\,.
\eea

\subsection{Matrix Elements}

\begin{figure}[t]
\includegraphics[width=0.48\textwidth,clip=true]{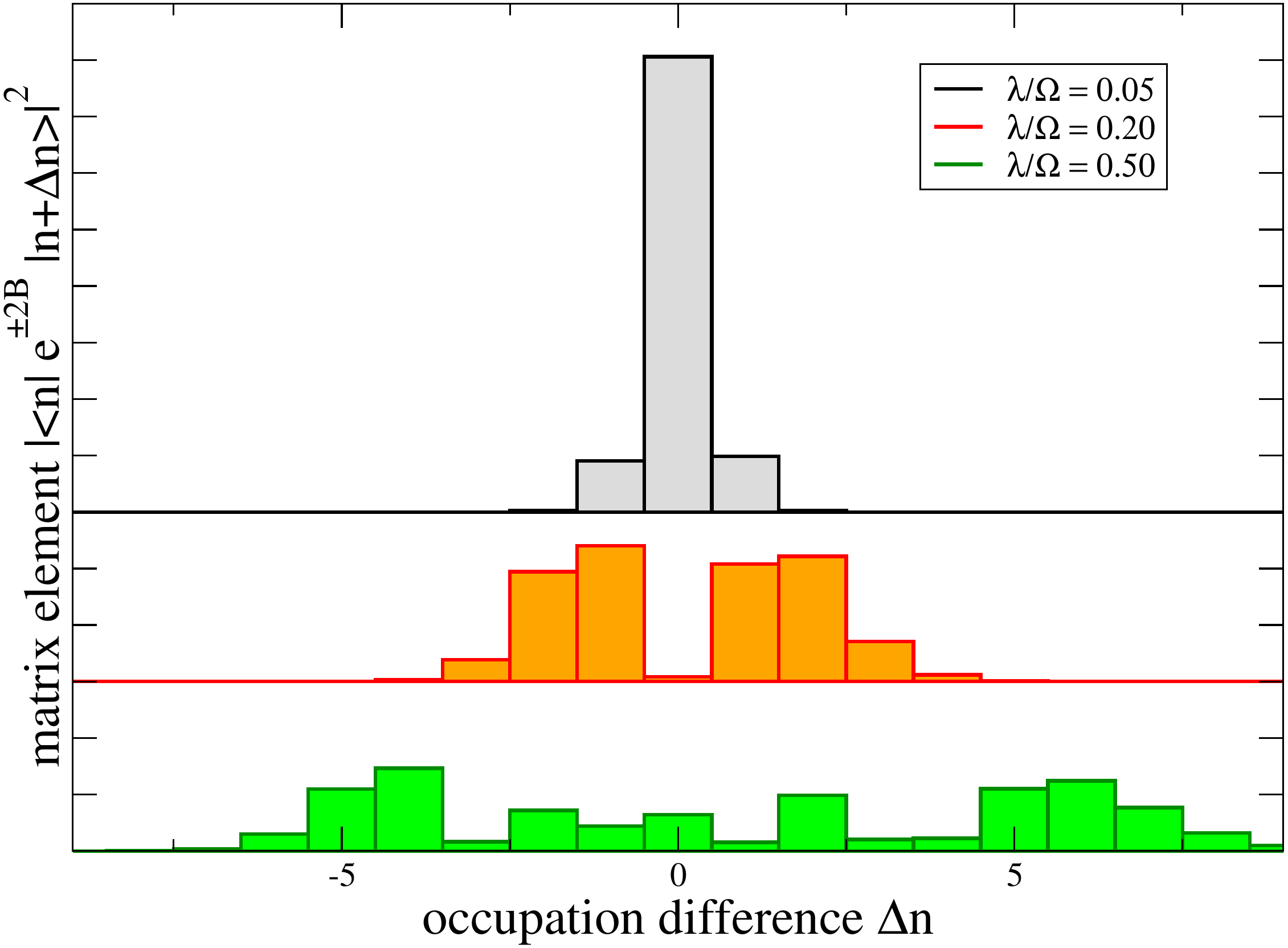}
\caption{\label{FIG:matrixel}(Color Online)
Matrix element $\abs{\widetilde{\bra{n}} e^{\pm 2 B} \widetilde{\ket{n+\Delta n}}}^2$ for different coupling strengths (vertically shifted for clarity).
For small $\lambda$ (top, black and grey), mainly transitions changing only the spin angular momentum quantum number are allowed.
As $\lambda$ increases (middle, red and orange and bottom, dark and light green), the matrix element also allows for 
transitions between distant occupation states.
All distributions are normalized to one, such that the plotted quantity can be interpreted as a conditional probability distribution. 
Tick marks on the vertical axis correspond to steps of $0.1$.
Other parameters: $n=10$.
}
\end{figure}

The matrix elements in the transition rates~(\ref{EQ:rate_general}) can also be conveniently evaluated using the polaron transformation.
For example, the collective spin flip operator -- recalling that $J_x = J_+ + J_-$ -- becomes
\bea\label{EQ:coupling1}
\abs{\bra{n,m} A_1 \ket{n'm'}}^2 &=& \abs{\widetilde{\bra{n,m}} U J_x U^\dagger \widetilde{\ket{n',m'}}}^2\nn
&=& \abs{\bra{\widetilde{n,m}}\left(J_+ e^{+2B} + {\rm h.c.} 
\right) \ket{\widetilde{n',m'}}}^2\nn
&=& \delta_{m',m-1} M_m^- \abs{\widetilde{\bra{n}} e^{+2 B} \widetilde{\ket{n'}}}^2\nn
&&+\delta_{m',m+1} M_m^+ \abs{\widetilde{\bra{n}} e^{-2 B} \widetilde{\ket{n'}}}^2\,,
\eea
and it is visible from the definition of $B$ in Eq.~(\ref{EQ:polaron_transformation}) that for finite $\lambda$ 
this reservoir triggers transitions between any $n$ and $n'$ but only
between neighboring $m$ and $m'=m\pm 1$.
Furthermore, it is straightforward to see that $\sum_n \abs{\widetilde{\bra{n}} e^{\pm 2 B} \widetilde{\ket{n'}}}^2 = \sum_{n'} \abs{\widetilde{\bra{n}} e^{\pm 2 B} \widetilde{\ket{n'}}}^2 = 1$.
Therefore, the absolute value of this matrix element can be interpreted as a conditional probability  
distribution spreading the original rate (for $\lambda=0$, admitting only $n'=n$) over different occupation eigenstates $n'$, see also Fig.~\ref{FIG:matrixel}.
Whereas for small $\lambda$ the matrix element is centered around $n'=n$, it becomes for large $\lambda$ more likely that also the bosonic occupation number changes.
We furthermore note the asymmetry of the distribution, which however is reduced when $\Delta n \ll n$. 
The fact that for stronger couplings the there is a dip at the origin can be qualitatively understood by realizing that $e^{\pm 2 B}$ is 
a displacement operator~\cite{scully1997}.

By contrast, the other coupling operators only allow for the creation or annihilation of one quantum of the longitudinal boson mode, respectively
\bea\label{EQ:coupling2}
\abs{\bra{n,m} A_2 \ket{n'm'}}^2 &=& \abs{\widetilde{\bra{n,m}} \left(a-\frac{\lambda^*}{\Omega} J_z\right) \widetilde{\ket{n'm'}}}^2\\
&=& \delta_{mm'} \delta_{n',n+1} (n+1)\nn
&& + \delta_{mm'} \delta_{nn'} \frac{4 m^2\lambda^2}{\Omega^2}\,,\nn
\abs{\bra{n,m} A_3 \ket{n'm'}}^2 &=& \abs{\widetilde{\bra{n,m}} \left(a^\dagger-\frac{\lambda}{\Omega} J_z\right) \widetilde{\ket{n'm'}}}^2\nn
&=& \delta_{mm'} \delta_{n',n-1} n + \delta_{mm'} \delta_{nn'} \frac{4 m^2\lambda^2}{\Omega^2}\,,\nonumber
\eea
Here, $\lambda$ only enters the diagonal contribution, which is irrelevant 
as it does not change the dynamics of the rate equation.

\begin{figure}[t]
\includegraphics[width=0.48\textwidth,clip=true]{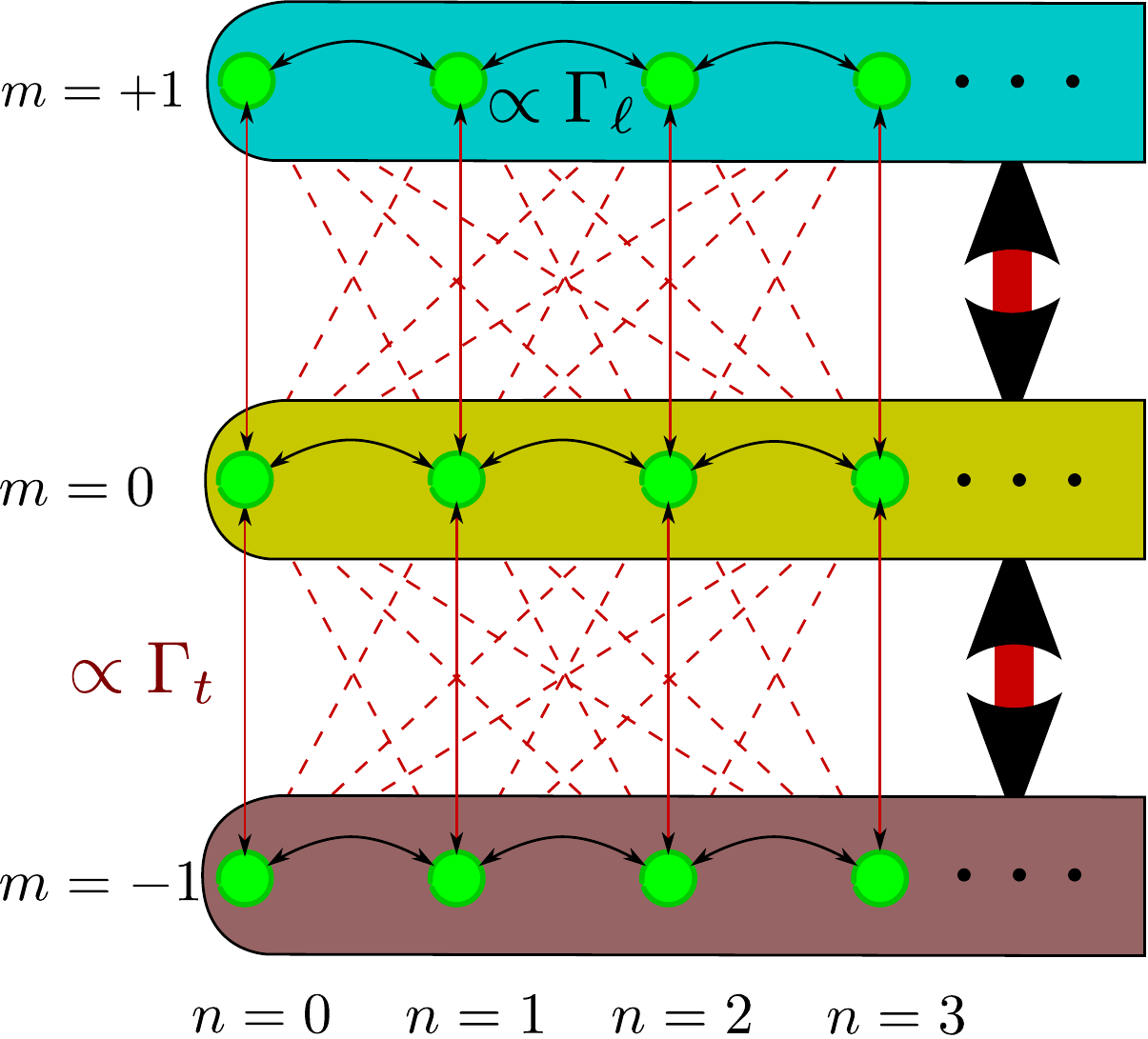}
\caption{\label{FIG:rategraph_sketch}(Color Online)
Graph representation of allowed transitions between the $n,m$-parametrized eigenstates (green circles) of $H_S$ for $N=2$.
Black arrows in horizontal direction represent transitions triggered by the longitudinal reservoir, compare Eq.~(\ref{EQ:coupling2}).
The other (red) lines represent transitions triggered by the transversal reservoir, compare Eq.~(\ref{EQ:coupling1}).
For finite $\lambda$, these also admit diagonal transitions changing both $m$ and $n$ (dashed red, background), whereas for $\lambda\to 0$, only the
vertical transitions (solid red) remain.
The basic idea of coarse-graining is to obtain effective rates (bold arrows) between meso-states (shaded regions) formed by lumping together the different 
Fock states for a given spin quantum number, physically motivated by fast horizontal equilibration $\Gamma_\ell \gg \Gamma_t$.
Whereas the transition rates in the original system obey local detailed balance relations~(\ref{EQ:localdetailedbalance}), 
the coarse-grained rates will not, cf. Eq.~(\ref{EQ:rate_coarsegrained}) and the discussion in Sec.~\ref{SEC:nonequilibrium_dynamics}.
}
\end{figure}

In total, the resulting rate equation 
\bea\label{EQ:full_rate_equation}
\dot P_{nm} = \sum_{n'm'} W_{nm,n'm'} P_{n'm'}
\eea
is of standard form, namely additive in the dissipators $W_{nm,n'm'} = W_{nm,n'm'}^{(t)}+W_{nm,n'm'}^{(\ell)}$, where for $(n,m)\neq(n',m')$ we have
for the (positive) transition rates -- cf. Eq.~(\ref{EQ:rate_general}) -- from ($n',m'$) to ($n,m$) the expressions 
\bea\label{EQ:full_transition_rates}
W_{nm,n'm'}^{(t)} &=& \gamma_{11}(E_{n'm'}-E_{nm}) \abs{\bra{nm} A_1 \ket{n'm'}}^2\,,\nn
W_{nm,n'm'}^{(\ell)} &=& \gamma_{23}(E_{n'm'}-E_{nm}) \abs{\bra{nm} A_3 \ket{n'm'}}^2\\
&&+\gamma_{32}(E_{n'm'}-E_{nm}) \abs{\bra{nm} A_2 \ket{n'm'}}^2\nonumber
\eea
The diagonal entries $W_{nm,nm}=-\sum\limits_{(n'm')\neq(nm)} W_{n'm',nm}$ follow from trace conservation.
We see that $\abs{\bra{n,m} A_1 \ket{n',m'}}^2 = \abs{\bra{n',m'} A_1 \ket{n,m}}^2$ and also that
$\abs{\bra{n,m} A_2 \ket{n'm'}}^2 = \abs{\bra{n'm'} A_3 \ket{n,m}}^2$, such that Eqns.~(\ref{EQ:ft_corrfunc}) imply
the usual local-detailed balance relations
\bea\label{EQ:localdetailedbalance}
\frac{W_{nm,n'm'}^{(t)}}{W_{n'm',nm}^{(t)}} &=& e^{\beta_t(E_{n'm'}-E_{nm})}\,,\nn
\frac{W_{nm,n'm'}^{(\ell)}}{W_{n'm',nm}^{(\ell)}} &=& e^{\beta_\ell(E_{n'm'}-E_{nm})}\,.
\eea
We note that this property enables one to formulate a consistent thermodynamic picture of these rate equations, including
positivity of the entropy production rate in a far-from-equilibrium regime ($\beta_t\neq \beta_\ell$) and the existence 
of a heat exchange fluctuation theorem~\cite{esposito2007a}.
The general structure of the rate equation is depicted in Fig.~\ref{FIG:rategraph_sketch}.

Any numerical simulation of the resulting rate equation~\cite{koch2005a} will have to cut the a priori infinitely large bosonic Hilbert space of the system.
With such a cutoff, the total dimension required by the rate equation scales as $N N_{\rm cut}$, but in particular for too small cutoff values, 
the dynamics can be altered as e.g. the bosonic commutation relations cannot be fulfilled at the cutoff.
One can in principle check for the influence of such a cutoff numerically by demonstrating convergence of results in dependence of the cutoff size $N_{\rm cut}$.
For steady state calculations the influence of the cutoff will be negligible when the highest considered occupation eigenstates are hardly populated.
In particular when one of the reservoir temperatures is large, one may however require a large boson cutoff $N_{\rm cut}$, making full-scale
simulations in this regime difficult.
It is therefore important to stress that coarse-graining procedure, that we introduce in Sec.~\ref{SEC:coarse_graining} below, leads
to an approximate description that does not require any cutoff, as all bosonic eigenstates are taken into account. 
We will then be able, within the range of validity of the coarse-grained picture, to obtain reliable numerical results also for large reservoir temperatures.

\subsection{Energy Currents}\label{SEC:currents}

When the rate matrix is additively decomposable into the reservoirs $\nu$, i.e., when the 
transition rates from energy eigenstate $j$ to energy eigenstate $i$ are decomposable as 
$W_{ij} = \sum_\nu W_{ij}^{(\nu)}$, 
we can directly infer the (time-dependent) energy currents from the system into reservoir $\nu$ by
multiplying the occupation $P_j$ with the reservoir-specific transition rate $W_{ij}^{(\nu)}$ and the
corresponding energy difference $(E_j-E_i)$.
Summing over all initial states and all allowed transitions then yields the energy current into
reservoir $\nu$
\bea\label{EQ:energycurrent_large}
I_E^{(\nu)}(t) &=& \sum_{i,j} (E_j - E_i) W_{ij}^{(\nu)} P_j(t)\nn
&\stackrel{t\to\infty}{\to}& \sum_{i,j} (E_j - E_i) W_{ij}^{(\nu)} \bar{P}_j\,.
\eea
Here, we have deliberately chosen the convention that the current counts positive when the system 
injects net energy into the reservoir, and negative otherwise.

\subsection{Coarse-Graining}\label{SEC:coarse_graining}

We can define the reduced probability of being in spin eigenstate $m$ by summing over the different occupation configurations
\bea
P_m = \sum_n P_{nm}\,.
\eea
Then, we can formally write its time derivative as
\bea
\dot{P}_m = \sum_{m'} \left[\sum_{nn'} W_{nm,n'm'} \frac{P_{n'm'}}{P_{m'}}\right] P_{m'}\,,
\eea
where we see that the set of $P_m$ does not obey a closed Markovian evolution equation, since to obtain the
time-dependent prefactors in square brackets one first has to solve the full rate equation for the $P_{nm}$ probabilities.
However, in certain limits an approximate Markovian description is possible.
To motivate this approximation we identify in the 
time-dependent prefactors in brackets the -- in general time-dependent -- conditional 
probability $P_{n'm'|m'}$ of the system being in state $n'm'$ provided that the 
spin is in state $m'$.
When parameters are now adjusted such that the longitudinal mode equilibrates much faster than the large spin, 
we can replace the time-dependent conditional probability by its stationary equilibrated value~\cite{esposito2012b}
\bea
\frac{P_{n'm'}}{P_{m'}} \to \bar{P}_{n'm'|m'}\,,
\eea
which in general depends on both states $n'$ and $m'$.
In general, the resulting coarse-grained rates
\bea
W_{mm'} &=& \sum_{nn'} W_{nm,n'm'} \bar{P}_{n'm'|m'}
\eea
between mesostates $m'$ and $m$ will implicitly depend on the coupling constants and temperatures of 
all reservoirs through the conditional steady-state probability $\bar{P}_{n'm'|m'}$.
This also holds for an additive decomposition of the total rate matrix
$W_{nm,n'm'} = \sum_\nu W_{nm,n'm'}^{(\nu)}$.
Therefore, in a coarse-grained description, the reservoirs will not simply enter as independent additive contributions, 
and furthermore, the coarse-grained rates need not obey detailed balance by construction
$\frac{W_{mm'}}{W_{m'm}} \neq e^{\beta(E_{m'}-E_{m})}$.

Specifically, we note that in our case the conditional probabilities  $\bar{P}_{n'm'|m'} = e^{-n' \beta_\ell \Omega}\left[1-e^{-\beta_\ell\Omega}\right]$
are just the thermalized probabilities of the longitudinal oscillator mode in contact with its own reservoir.
They are well approached when $\Gamma_\ell \gg \Gamma_t$.
The approximate coarse-grained rates only describe transitions between the large spin eigenstates
\bea
W_{mm'} &=& \sum_{nn'} W_{nm,n'm'} e^{-n' \beta_\ell \Omega}\left[1-e^{-\beta_\ell\Omega}\right]\,.
\eea
We note that the rates due to $A_2$ and $A_3$ will not contribute to the coarse-grained ones $W_{mm'}^{(\ell)} = 0$, 
since they do not induce transitions between different mesostates $m$ and $m'$, see Eq.~(\ref{EQ:coupling2})
and Fig.~\ref{FIG:rategraph_sketch}.
For the other rates however, a contribution remains, such that we can write
\bea\label{EQ:rate_coarsegrained}
W_{mm'} &=& \delta_{m',m-1} M_m^- \sum_{nn'} \gamma_{11}(E_{n'm'}-E_{nm})\times\nn
&&\times \abs{\widetilde{\bra{n}} e^{+2 B} \widetilde{\ket{n'}}}^2 e^{-n' \beta_\ell \Omega}\left[ 1-e^{-\beta_\ell \Omega }\right]\nn
&&+\delta_{m',m+1} M_m^+ \sum_{nn'} \gamma_{11}(E_{n'm'}-E_{nm}) \times\nn
&&\times \abs{\widetilde{\bra{n}} e^{-2 B} \widetilde{\ket{n'}}}^2 e^{-n' \beta_\ell \Omega}\left[ 1-e^{-\beta_\ell \Omega }\right]\,,
\eea
where we have used Eq.~(\ref{EQ:coupling1}).
We stress again that, unless the temperatures of both reservoirs are equal, the coarse-grained rates will not obey a conventional 
detailed balance relation.
Instead, we find a more general relation, see Sec.~\ref{SEC:nonequilibrium_dynamics}.
With introducing the net number of bosons exchanged with the longitudinal boson reservoir $\bar{n} = n-n'$ 
(we will in the following use the overbar to indicate that $\bar{n}$ can assume negative values) and defining the
coarse-grained spin energy as
\bea\label{EQ:spectrum_spin}
E_m = \omega_0 m - 4 \frac{\abs{\lambda}^2}{\Omega} m^2\,,
\eea
we can rewrite the approximate coarse-grained rates as
\bea
W_{mm'} &=& \delta_{m',m-1} M_m^- \sum_{\bar{n}=-\infty}^{+\infty} \gamma_{11}(E_{m-1}-E_m-\Omega \bar{n}) \alpha_{\bar{n}}\qquad\nn
&&+\delta_{m',m+1} M_m^+ \sum_{\bar{n}=-\infty}^{+\infty} \gamma_{11}(E_{m+1}-E_m - \Omega\bar{n}) \alpha_{\bar{n}}\,.\nn
\eea
Above, we have introduced a normalized distribution with $\sum_{\bar{n}=-\infty}^{+\infty} \alpha_{\bar{n}} = 1$ by using
\bea
\alpha_{\bar{n}} &=& \sum_{nn'=0}^\infty \delta_{\bar{n},n-n'} \abs{\widetilde{\bra{n}} e^{\pm 2B} \widetilde{\ket{n'}}}^2 
e^{-n' \beta_\ell \Omega}\left[1-e^{-\beta_\ell \Omega }\right]\nn
&=& e^{-\frac{4\abs{\lambda}^2}{\Omega^2} (1 + 2 n_\ell)} 
\left(\frac{1+n_\ell}{n_\ell}\right)^{\bar{n}/2}\times\nn
&&\times {\cal J}_{\bar{n}}\left(\frac{8\abs{\lambda}^2}{\Omega^2} \sqrt{n_\ell(1+n_\ell)}\right)\,,
\eea
where  ${\cal J}_{\bar{n}}(x)$ denotes the modified Bessel function of the first kind.
In the second line, we have explicitly evaluated the matrix element and performed the summation.

\begin{figure}[t]
\includegraphics[width=0.45\textwidth,clip=true]{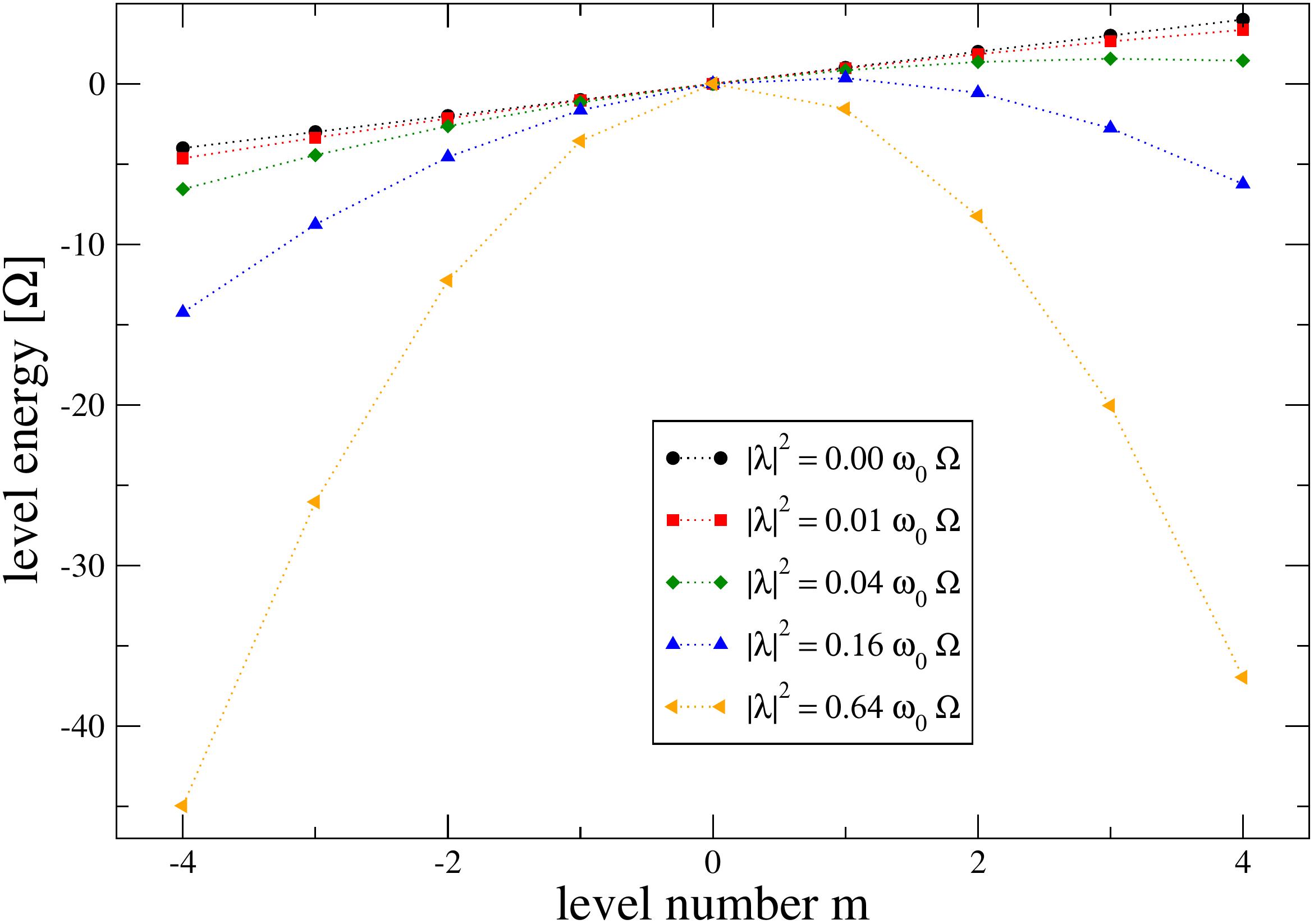}
\caption{\label{FIG:excitation}
Level spectrum~(\ref{EQ:spectrum_spin}) for $N=8$.
For finite coupling strength, the level spectrum (symbols) is no longer equidistant.
Since a transition from the $m=+N/2$ state (right-most) to the ground state $m=-N/2$ (left-most) is only allowed along the connected states (dotted lines), 
the level renormalization may seriously affect the Dicke superradiance and even block the process.
Other parameters: $\omega_0=\Omega$.
}
\end{figure}

The coarse-grained rates depend on both $\beta_\ell$ (through $\alpha_{\bar{n}}$) and $\beta_t$ (through $\gamma_{11}(\omega)$).
Despite these sophisticated rates, the structure of the approximate coarse-grained rate equation now has a 
simple tri-diagonal form
\bea\label{EQ:modified_rate_equation}
\dot P_m &=&-\left[\gamma(E_m-E_{m-1}) M_m^- + \gamma(E_m-E_{m+1}) M_m^+\right] P_m\nn
&&+\gamma(E_{m+1}-E_m) M_m^+ P_{m+1}\nn
&&+\gamma(E_{m-1}-E_m) M_m^- P_{m-1}\,,\nn
\gamma(\omega) &=& \sum_{\bar{n}=-\infty}^{+\infty} \gamma_{\bar{n}}(\omega) 
= \sum_{\bar{n}=-\infty}^{+\infty} \gamma_{11}(\omega - \Omega \bar{n}) \alpha_{\bar{n}}
\eea
with dimension $N+1$.
We see that in contrast to the original superradiance master equation~(\ref{EQ:rate_equation}), 
the level spectrum~(\ref{EQ:spectrum_spin}) is no longer equidistant, but the scaling of the matrix elements 
with the spin length $N$ will persist.
This also implies that the most excited spin state is not necessarily the one with $m=+N/2$, see Fig.\ref{FIG:excitation}.

Although it is only valid in the limit where the longitudinal mode equilibrates much faster than the large spin ($\Gamma_\ell \gg \Gamma_t$), the advantages of the coarse-graining
procedure are obvious.
The numerical effort is significantly reduced as the rate matrix describing both the large spin and the longitudinal phonon mode has dimension $\ord\{N N_{\rm cut}\}$
and is fully occupied, whereas the coarse-grained rate matrix is only $N$-dimensional and tri-diagonal.
Furthermore, a bosonic cutoff is not necessary in the coarse-grained description and a thermodynamic interpretation is still possible, provided that the 
generalized KMS relations~(\ref{EQ:kms_relation}) are correctly taken into account.

\subsection{Non-equilibrium dynamics}\label{SEC:nonequilibrium_dynamics}

The temperatures of transversal and longitudinal reservoirs 
may be different, giving rise to non-equilibrium stationary energy currents between the reservoirs.
To evaluate the total energy exchanged with both reservoirs one could of course numerically evaluate the high-dimensional
rate equation with a suitable cutoff of the longitudinal boson mode.
Especially at large longitudinal temperatures, however, this may be computationally difficult.

To obtain the thermodynamics of the coarse-grained rate equation, it is helpful to note that the 
standard Kubo-Martin-Schwinger (KMS) relations of transversal and longitudinal 
correlation functions imply a non-standard~\cite{schaller2013a,krause2015a}
KMS-type relation between the different terms in the sum
\be\label{EQ:kms_relation}
\frac{\gamma_{+\bar{n}}(-\omega)}{\gamma_{-\bar{n}}(+\omega)} = \frac{\alpha_{+\bar{n}}}{\alpha_{-\bar{n}}} \frac{\gamma_{11}(-\omega-\Omega \bar{n})}{\gamma_{11}(+\omega+\Omega \bar{n})}
= e^{+\beta_\ell \bar{n} \Omega} e^{-\beta_t(\omega+\bar{n}\Omega)}\,.\qquad
\ee
Naturally, at equilibrium $\beta_\ell = \beta_t$ we recover the usual KMS relation, the coarse-grained rates obey detailed balance, 
and the stationary state of the coarse-grained rate equation~(\ref{EQ:modified_rate_equation}) is just the canonical equilibrium one.

For different temperatures, relation~(\ref{EQ:kms_relation}) is consistent with the interpretation that 
processes described by the term $\gamma_{\bar n}(\omega)$ -- representing a net change of $-\omega$ in the system's energy via the exchange
of $\bar{n}$ bosons with the longitudinal mode -- must be accompanied with an energy transfer of $\bar{n} \Omega$ into the longitudinal reservoir and a transfer
of $\omega - \bar{n} \Omega$ into the the transversal reservoir.
We can quantify for each
process the fractions of the system's energy change that are transferred into longitudinal and transversal reservoirs, respectively.
Formally, we can then compute for a rate equation of the form 
$\dot{P}_i = \sum_j \sum_{\bar{n}} W_{ij}^{(\bar{n})} P_j$ the stationary energy currents into both reservoirs via
\bea
I_E^{(\ell/t)} &=& \sum_{i,j} \sum_{\bar{n}} \Delta E_{ij}^{(\bar{n},\ell/t)} W_{ij}^{(\bar{n})} \bar{P}_j\,.
\eea
Specifically, we do this for each transition term in the rate equation~(\ref{EQ:modified_rate_equation}) 
\bea\label{EQ:mod_rate_equation_cf}
\dot P_m &=&-\left[\gamma(E_m-E_{m-1}) M_m^- + \gamma(E_m-E_{m+1}) M_m^+\right] P_m\nn
&&+\sum_{\bar{n}} \gamma_{\bar{n}}(E_{m+1}-E_m) M_m^+ P_{m+1}
\nn
&&+\sum_{\bar{n}} \gamma_{\bar{n}} (E_{m-1}-E_m) M_m^- P_{m-1}
\,,
\eea
to calculate the currents into reservoirs $\ell$ and $t$ via
\bea\label{EQ:current_explicit}
I_E^\ell(t) &=& \sum_m \sum_{\bar{n}} \bar{n} \Omega \gamma_{\bar{n}} (E_{m+1}-E_m) M_m^+ P_{m+1}(t)\nn
&&+ \sum_m \sum_{\bar{n}} \bar{n} \Omega \gamma_{\bar{n}} (E_{m-1}-E_m) M_m^- P_{m-1}(t)\nn
I_E^t(t) &=& \sum_m \sum_{\bar{n}} (E_{m+1}-E_m-\bar{n}\Omega) \times\nn
&&\times\gamma_{\bar{n}} (E_{m+1}-E_m) M_m^+ P_{m+1}(t)\nn
&&+ \sum_m \sum_{\bar{n}} (E_{m-1}-E_m-\bar{n} \Omega)\times\nn
&&\times\gamma_{\bar{n}} (E_{m-1}-E_m) M_m^- P_{m-1}(t)\,.
\eea

Before proceeding, we mention a few properties of the steady-state currents $I_E^{\ell/t} = \lim\limits_{t\to\infty} I_E^{\ell/t}(t)$, 
obtained by letting $P_m(t)\to \bar{P}_m$.
Firstly, in equilibrium ($\beta_\ell=\beta_t=\beta$), both currents must vanish individually.
Formally, this is enforced by the KMS relation~(\ref{EQ:kms_relation}).
Secondly, the steady-state currents must compensate as the total energy is conserved $I_E^\ell = - I_E^t$.
Thirdly, the second law of thermodynamics actually implies that $(\beta_\ell - \beta_t) I_E^\ell \ge 0$ in all parameter regimes (heat always flows from hot to cold).
Finally, we also stress that a necessary (but not sufficient) condition for a finite current is that all stationary probabilities must be strictly
smaller than one (if only one state is occupied $P_{\bar{m}}=1$, there are no transitions between two states and thus no stationary current).

In our results section, we explicitly confirm that the currents obtained from the coarse-grained rate equation~(\ref{EQ:mod_rate_equation_cf}) and from the
high-dimensional rate equation~(\ref{EQ:full_rate_equation}) coincide in the appropriate limit $\Gamma_\ell \gg \Gamma_t$.

Finally, to perform calculations, we will parametrize the spectral coupling density in Eq.~(\ref{EQ:ft_corrfunc}) 
with an ohmic form and exponential cutoff $\omega_c$~\cite{brandes2005a}
\bea\label{EQ:spec_coup_dens}
\Gamma_t(\omega) = \Gamma_t \omega e^{-\omega/\omega_c}\,.
\eea
Here, $\Gamma_t$ regulates the coupling strength to the transversal reservoir and the cutoff expresses the fact 
that for any realistic model the spectral coupling density should decay in the ultraviolet regime.


\subsection{Weak-Coupling limit}

The observation that the two reservoirs no longer enter additively in the coarse-grained description 
does not come unexpected, as an additive decomposition typically requires the weak-coupling limit
between system and reservoir.
By contrast, in our setup the large spin may be strongly coupled to the longitudinal
boson mode.
To check for consistency, we will therefore briefly discuss the limit of small $\lambda$.
We can use the fact that near the origin we have ${\cal J}_{\bar{n}}(x) = \frac{x^{\abs{\bar{n}}}}{2^{\abs{\bar{n}}} \abs{\bar{n}}!} + \ord\{x^{\abs{\bar{n}}+2}\}$
to expand the correlation function in the dissipator as
\bea
\gamma(\omega) &=& \gamma_{11}(\omega) + \frac{4\abs{\lambda}^2}{\Omega^2} \Big[\gamma_{11}(\omega+\Omega) n_\ell\nn
&&\;\; + \gamma_{11}(\omega-\Omega)(1+n_\ell) - \gamma_{11}(\omega) (1+2n_\ell)\Big]\nn
&&+ \ord\{\abs{\lambda}^4\}\,.
\eea
Clearly, the original Dicke superradiance model is consistently recovered at $\lambda\to 0$.
As expected, we obtain an additional dissipator of order $\abs{\lambda}^2$.
What at first sight comes a bit unexpected is that the additional dissipator does not solely depend on the thermal
properties of reservoir $\ell$ but also on reservoir $t$ (through $\gamma_{11}$).
It is also proportional to the product of $\Gamma_t (\omega)$ and $\abs{\lambda}^2$.
This however is fully consistent with our initial model, since the interaction mediated by the $\lambda$-coupling is 
of pure-dephasing type for the large spin.
Therefore, if applied alone ($\Gamma_t(\omega) \to 0$), it should not affect the dynamics of the angular momentum 
eigenstates at all but can only induce dephasing of coherences between different angular momentum eigenstates.

\section{Results}

\subsection{Equilibrium: Superradiant decay}\label{SEC:decay}

For vanishing coupling $\lambda=0$, it is well-known that at zero temperature, the quadratic scaling of the matrix elements leads to
superradiant decay toward the ground state with a maximum intensity scaling as $N^2$ and consequently a width of the peak scaling as $1/N$.
We can reproduce these findings in the appropriate limits (not shown).
In this section, we want to investigate how the decay dynamics is influenced by the presence of the longitudinal mode and therefore
consider the case that both reservoirs are at zero temperature, or at least temperatures sufficiently low that excitations entering the 
system from the reservoir can be safely neglected,  $n_\ell=n_t=0$.

The scaling of Dicke superradiance also reflects in the passage time towards the ground state.
At zero temperature, the probability distribution for the passage time to the ground state is defined by
\bea
P(t) = \frac{d}{dt} P_{-N/2}(t)\,.
\eea
Provided the ground state is the stationary state associated with a given initial state, 
it is straightforward to check that it is normalized $\int_0^\infty P(t) dt = 1$ and positive $P(t) \ge 0$.
We will be interested in the mean passage time and its width, which requires us to evaluate 
\bea
\expval{\tau^n} &=& \int_0^\infty \tau^n P(\tau) d\tau
\eea
for $n=1$ and $n=2$.
For a rate matrix of the form
\bea
{\cal L} = \left(\begin{array}{cccc}
0 &  L_{12} &\\
0 & -L_{12} &  L_{23}\\
  &         & -L_{23} & \ddots\\
  &         &         & \ddots
\end{array}\right)\,,
\eea
the first and second cumulants of the passage time distribution assume the simple form
\bea
\expval{\tau} &=& \frac{1}{L_{12}} + \frac{1}{L_{23}} + \ldots\,,\nn
\expval{\tau^2}-\expval{\tau}^2 &=& \frac{1}{L_{12}^2} + \frac{1}{L_{23}^2} + \ldots\,.
\eea
Without longitudinal boson coupling $\lambda=0$, we have 
$L_{12}=\Gamma_t(\omega_0) M_{-N/2}^+$, 
$L_{23}=\Gamma_t(\omega_0) M_{-N/2+1}^+$ and so on -- cf.~Eq.~(\ref{EQ:rate_equation}) --  
such that we can obtain the mean passage time and its width for the original Dicke limit analytically
\bea
\expval{\tau} &=& \frac{2\left(\gamma+\Psi_0(N+1)\right)}{\Gamma_t(\omega_0) (N+1)}\,,\\
\expval{\tau^2}-\expval{\tau}^2 &=& \Big[12\gamma + \pi^2(N+1) + 12 \Psi_0(N+1)\nn
&& - 6 (N+1) \Psi_1(N+1)\Big]/\left[3 \Gamma_t^2(\omega_0) (N+1)^3\right]\,,\nonumber
\eea
where $\gamma\approx 0.577216$ denotes the Euler constant and $\Psi_n(x)$ the Polygamma function.
We see that for large $N$ the mean passage time roughly scales as 
$\expval{\tau} \approx (2\gamma + \ln N)/(\Gamma_t(\omega_0) N)$ and the width as
$\sqrt{\expval{\tau^2}-\expval{\tau}^2} \approx \pi/(\sqrt{3} N \Gamma_t(\omega_0))$.
That means that to obtain a sharply determined passage time one requires very large $N$, e.g.\ to obtain a width ten times smaller
than the mean one requires $N=\ord\{10^7\}$ two-level systems.
For infinite $N$, the passage time is very well determined and -- despite the stochastic nature of the rate equation -- the system 
relaxes nearly deterministically towards the ground state with a negligible temporal error.

\begin{figure}[t]
\includegraphics[width=0.48\textwidth,clip=true]{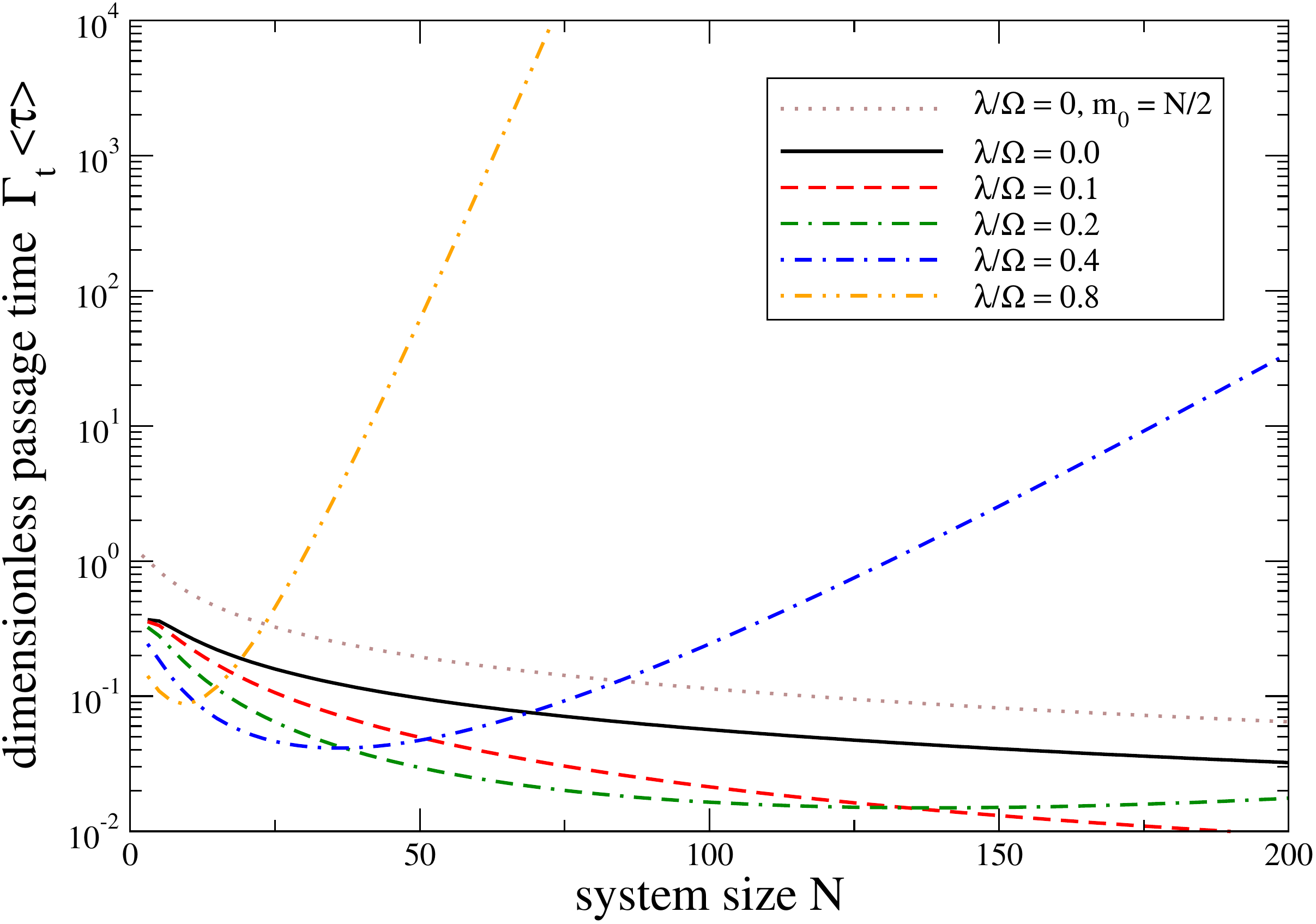}
\caption{\label{FIG:passagetime}(Color Online)
Plot of the dimensionless mean passage time versus system size $N$ for different coupling strengths when both reservoirs are at the same low temperature $\beta=\beta_\ell=\beta_t$.
The dynamics of the original superradiance scenario (dotted brown) is hardly changed when instead from starting at $m=+N/2$ we do initially prepare the
system at $m=-1/2$ (solid black) -- apart from an obvious speedup by a factor two.
When the coupling strength $\lambda$ is increased, the presence of additional decay channels (compare the dashed lines in Fig.~\ref{FIG:rategraph_sketch}) 
first increases the relaxation speed (dashed red and dash-dash-dotted green).
However, beyond a critical system size $N$ an exponential slowdown of relaxation (increase of the passage time) occurs (dash-dotted blue and dash-dot-dotted orange).
This is due to the renormalization-induced increase of the excitation energies above the ground state (compare e.g.\ the orange curve in Fig.~\ref{FIG:excitation}), 
which due to the finite bandwidth $\omega_c$ finds no support in the correlation function.
Other parameters: $\omega_0=\Omega$, $\beta\Omega = 10$, $\omega_c = 10 \Omega$.
}
\end{figure}

These findings would be qualitatively similar if we start from the middle of the spectrum (e.g.\ at $m\approx0$) instead.
In fact, to investigate how the additional boson mode influences the relaxation behaviour to the ground state at low temperatures we have
to take the level distortion in Fig.~\ref{FIG:excitation} into account.
When preparing the system in the state $m=+N/2$ we may not see any relaxation toward the ground state as a trivial effect of the level 
renormalization.
To ensure that we only observe unidirectional relaxation we therefore constrain ourselves to odd $N$ and prepare the system initially in the
state $m=-1/2$.
The results are displayed in Fig.~\ref{FIG:passagetime}.
One can see that at first finite couplings $\lambda$ aid the relaxation process, since the passage time becomes shorter.
However, above a critical coupling strength the passage time increases again for larger system sizes $N$.
This is due to the finite bandwidth $\omega_{\rm c}$ of the spectral coupling density~(\ref{EQ:spec_coup_dens}): 
The excitation energy above the ground state $\Delta E = \omega_0 + 4 \frac{\abs{\lambda}^2}{\Omega}(N-1)$ 
becomes so large $\abs{\Delta E} \gg \omega_c$ that the bosonic correlation function has no support and the last steps above the ground state occur extremely slow, 
with the visible effect on the passage time.

\subsection{Non-Equilibrium: Steady state heat current and rectification}\label{SEC:current}

When we consider different temperatures in both reservoirs, this will induce a steady state heat current from hot to cold across the system.
This simply means that trajectories where the system absorbs energy from the hot reservoir and afterwards emits energy into the cold reservoir 
become more likely than trajectories where the net flow of energy is opposed.
The total energy is conserved, which at steady state implies that we need to consider only the energy current into the 
longitudinal boson reservoir $I_E = I_E^\ell= -I_E^t$ (we have of course confirmed this equality).
As heat currents are driven by transitions between energy eigenstates, this means that the obtain a non-vanishing current, 
the system should not reside in a pure state.
Eqns.~(\ref{EQ:energycurrent_large}) and~(\ref{EQ:current_explicit}) imply that to calculate a current, we first have to evaluate the stationary probabilities, which can for
large matrices be numerically unstable.
Therefore, we provide an analytical formula for tri-diagonal rate matrices in Appendix~\ref{APP:stationary_state}.

\subsubsection{Weak-coupling Current}

We parametrize the inverse temperatures as
\bea\label{EQ:betafunc}
\beta_\ell &=& \frac{1}{2}\left[\bar\beta+\Delta\beta+\sqrt{\bar\beta^2+\Delta\beta^2}\right]\,,\nn
\beta_t &=& \frac{1}{2}\left[\bar\beta-\Delta\beta+\sqrt{\bar\beta^2+\Delta\beta^2}\right]
\eea
and plot the energy current through the system versus $\Delta\beta$ for a fixed inverse average temperature $\bar\beta$.
Trivially, as a consequence of the second law we expect for $\Delta\beta>0$ (implying for the temperatures $T_\ell < T_t$) that
the current entering the longitudinal reservoir is positive $I_E > 0$ and for $\Delta\beta<0$ we consequently expect $I_E < 0$.

\begin{figure}[t]
\includegraphics[width=0.48\textwidth,clip=true]{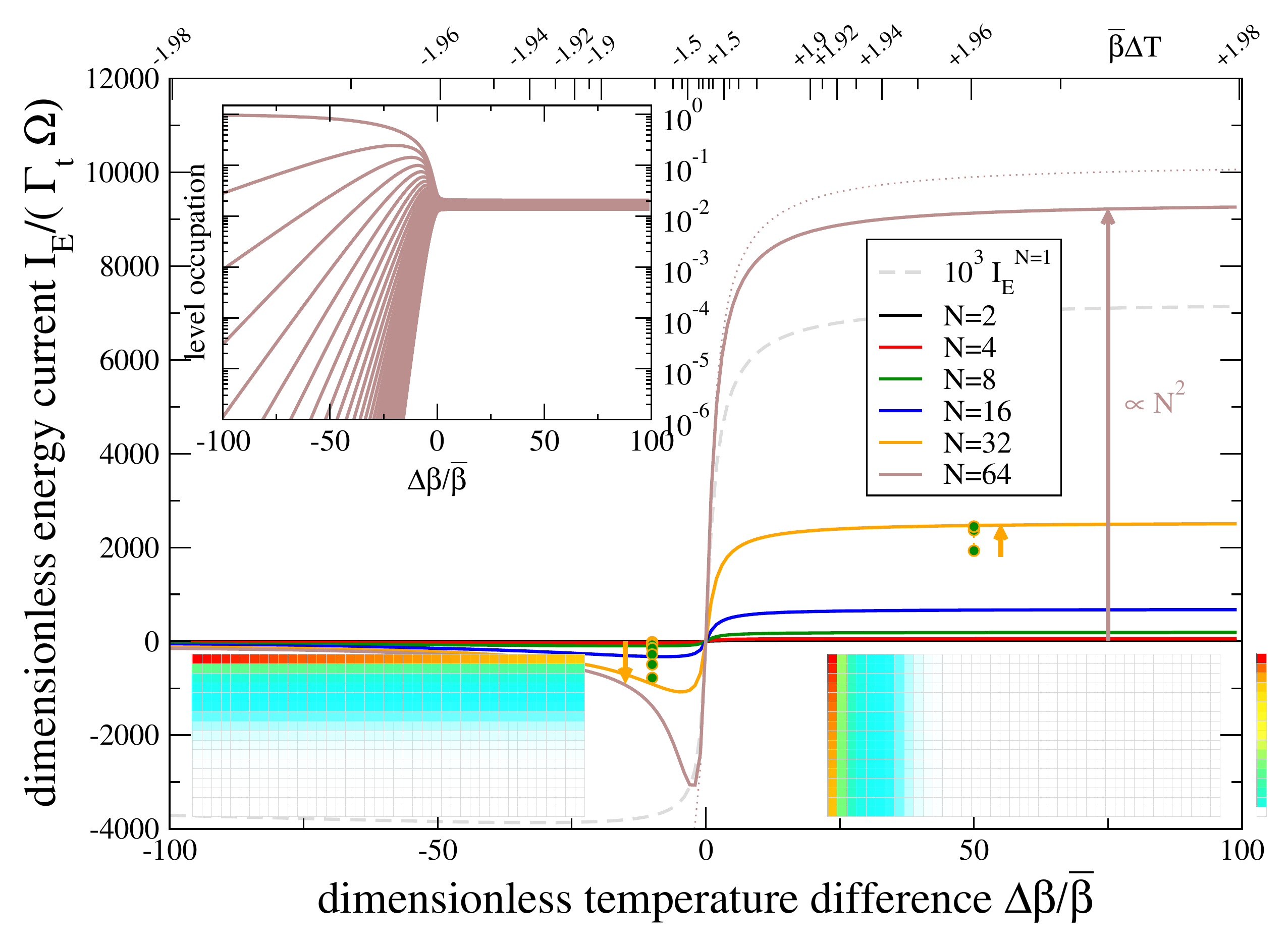}
\caption{\label{FIG:energycurrent}(Color Online)
Plot of the dimensionless steady state energy current versus dimensionless inverse temperature difference 
for weak coupling $\lambda=0.1\Omega$ and large average temperatures $\bar\beta\Omega=0.01$.
The top horizontal axis converts into dimensionless temperature differences $\bar\beta\Delta T = \bar\beta(T_t-T_\ell)$.
The top left inset shows the stationary occupation of the energy levels for $N=64$, where the top curve denotes the ground state and
the bottom curve the most excited state (at $\Delta\beta/\bar\beta=0$ we have the Gibbs distribution at $\bar\beta$).
Finally, the two density plots (white corresponds to zero, red to respective maximum) display the 
stationary state occupation $\bar{P}_{nm}$ of the full rate equation~(\ref{EQ:full_rate_equation}) 
for $N=16$, $N_{\rm cut} = 40$, and $\Gamma_\ell = 10^6 \Gamma_t$ at $\Delta\beta/\bar\beta=-100$ (left) 
and $\Delta\beta/\bar\beta=+100$ (right).
For the density plots, $m$ ranges from $-N/2$ (top) to $+N/2$ (bottom), $n$ ranges from $0$ (left) to $N_{\rm cut}$ (right).
When the levels are approximately equally occupied ($\Delta\beta/\bar\beta \stackrel{>}{\approx} 0$, and right density plot), the 
quadratic scaling of the matrix elements around $m=0$ carries over to the stationary current
as predicted in Eq.~(\ref{EQ:current_approx}) (thin dotted line for $N=64$).
In contrast, for $\Delta\beta/\bar\beta \ll 0$, the system dominantly resides in the lowest mesostate (inset and left density plot), 
and the current is consequently strongly suppressed.
In total, the system may therefore be used as a heat diode. 
In contrast, the current for $N=1$ (dashed grey, scaled by $10^3$ for visibility) is also asymmetric but does not display significant rectification.
The symbols (for $N=32$ only) indicate currents derived using the full rate equation~(\ref{EQ:full_rate_equation}) 
with a maximum occupation of the longitudinal boson mode $N_{\rm cut}\in\{5,10,20,40,80,160,320\}$ (orange arrows)
and $\Gamma_\ell = 10^6 \Gamma_t$.
With increasing cutoff $N_{\rm cut}$, the full master equation current (symbols) approaches the coarse-grained one 
(orange curve), where convergence is much faster when the longitudinal temperature is low:
For large longitudinal temperatures $T_\ell$ (left), 320 bosonic modes barely suffice to ensure convergence, 
whereas for small temperatures $T_\ell$ (right), roughly 10 modes suffice.
Other parameters: $\bar\beta\Gamma_t=0.01$, $\omega_0=\Omega$, $\lambda=0.1\Omega$, $\omega_{\rm c} = 10 \Omega$.
}
\end{figure}

To drive the system into a regime where the stationary current is mainly carried by large matrix elements $M_m^\pm$ and thus 
scales quadratically with the size $N$, we essentially have to populate the states with $m\approx 0$ as these
contribute most to the current, compare Eq.~(\ref{EQ:current_explicit}).
For our model, such a configuration is best approached when all populations are approximately equally occupied:
For weak coupling strengths $\lambda$ we can approximate $E_{m+1}-E_m\approx\omega_0$ such that the summation in the 
current from the equipartition assumption simply
yields a quadratic factor $\sum_m M_m^\pm \bar P_{m\pm 1} = N(N+2)/6$.
From Eq.(\ref{EQ:current_explicit}) we then obtain that the current will scale quadratically with $N$ in this regime
\bea\label{EQ:current_approx}
I_E \to \sum_{\bar{n}} (\bar{n} \Omega)\left[\gamma_{\bar{n}}(+\omega_0)+\gamma_{\bar{n}}(-\omega_0)\right] \frac{N(N+2)}{6}\,.
\eea
Such an equipartition regime can be expected at large average temperatures, and to see a significant current we do at the
same time require a large temperature difference.
Transferred to our variables in Eq.~(\ref{EQ:betafunc}) this means we have to consider small $\bar\beta$ and large $\Delta \beta$.
Fig.~\ref{FIG:energycurrent} indeed shows a quadratic scaling of the current with $N$ in the regime where
the populations are approximately equal (positive $\Delta\beta$).
The thin dotted line for $N=64$ also demonstrates the quality of the analytic approximation~(\ref{EQ:current_approx}).
Most interesting however, we observe that for large temperature differences, under temperature inversion ($\Delta\beta\to-\Delta\beta$) 
the populations are no longer equally occupied and simultaneously the absolute value of the current drops drastically.
This is related to a {\em configurational blockade}, where the coarse-grained system relaxes to a pure state (see caption of Fig.~\ref{FIG:energycurrent}).
Thus, at large temperature differences the system effectively 
implements a heat diode~\cite{segal2008b,ojanen2009a,ruokola2011a} with a rectification efficiency 
that is controllable by $N$.
Since for $\Delta\beta \gg 0$ the current scales quadratically and for $\Delta\beta \ll 0$ it does not, 
this effect can be controlled by increasing the number of two-level systems $N$.
For a small negative thermal bias the occupations of higher levels drop only mildly (implying a finite current) whereas 
for $\Delta\beta \ll 0$ essentially just the ground state is occupied (inset).
This also results in a negative differential thermo-conductance~\cite{shao2011a,hu2011a,ren2013a,sierra2014a}.

Finally, we would like to stress that we can compare the current from the coarse-grained rate equation~(\ref{EQ:current_explicit}) with the
one computed from the exact master equation when $\Gamma_\ell \gg \Gamma_t$. 
This requires to take a sufficient number of maximum bosonic occupations into account, requiring potentially large computational resources.
The symbols for $N_{\rm cut}\in\{5,10,20,40,80,160,320\}$ in Fig.~\ref{FIG:energycurrent} demonstrate that convergence for the current is reached in either regime
(also demonstrating validity of the coarse-graining approximation), but it is significantly 
slower when the temperature of the longitudinal boson reservoir is large (negative $\Delta \beta$).
This is somewhat expected, since for large temperatures many longitudinal boson mode excitations have to be taken into account.
Consistently, we see in the bottom left density plot of Fig.~\ref{FIG:energycurrent} that the occupation for the state $\ket{-N/2, +N_{\rm cut}}$ is 
not negligible for the chosen cutoff value $N_{\rm cut}=40$.

\begin{figure}[t]
\includegraphics[width=0.48\textwidth,clip=true]{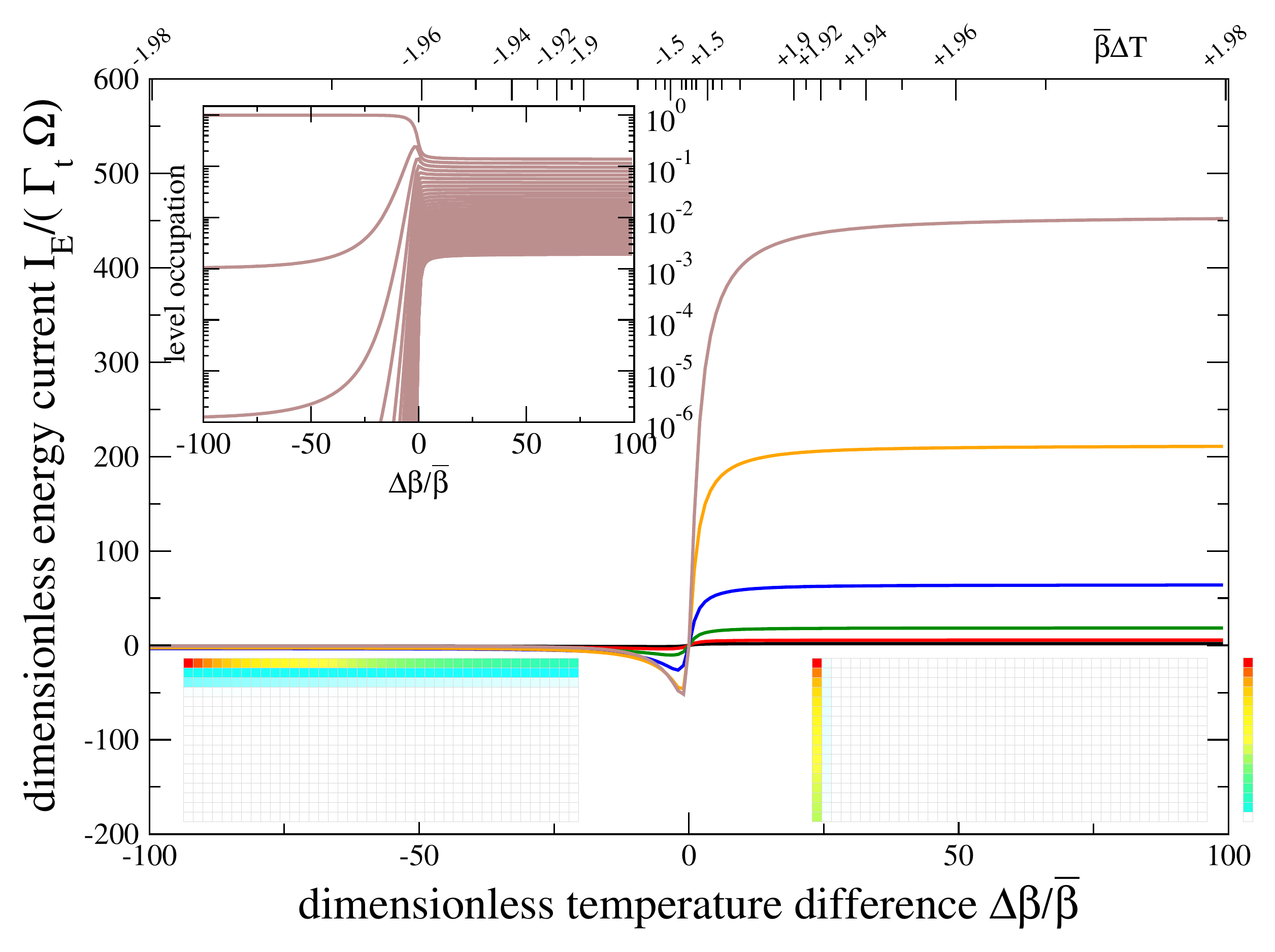}
\caption{\label{FIG:energycurrenta}(Color Online)
Similar as Fig.~\ref{FIG:energycurrent}, but for a lower average temperature $\bar\beta\Omega=0.1$.
An equipartition of levels is not reached, and the current does not scale quadratically with $N$.
Nevertheless, rectification is still present.
Color codes and other parameters have been chosen as in Fig.~\ref{FIG:energycurrent}.
}
\end{figure}

The maximum bosonic cutoff can be reduced when one lowers the average temperature.
Indeed, we see in Fig.~\ref{FIG:energycurrenta} that for $N=16$ (left density plot) fewer bosonic modes are occupied.
However, the reduction of the average temperature (increase of $\bar\beta$) also has the effect that the
levels in the conducting direction are no longer equipartitioned, such that the current is reduced and does no longer
scale quadratically in $N$.

\begin{figure}[t]
\includegraphics[width=0.48\textwidth,clip=true]{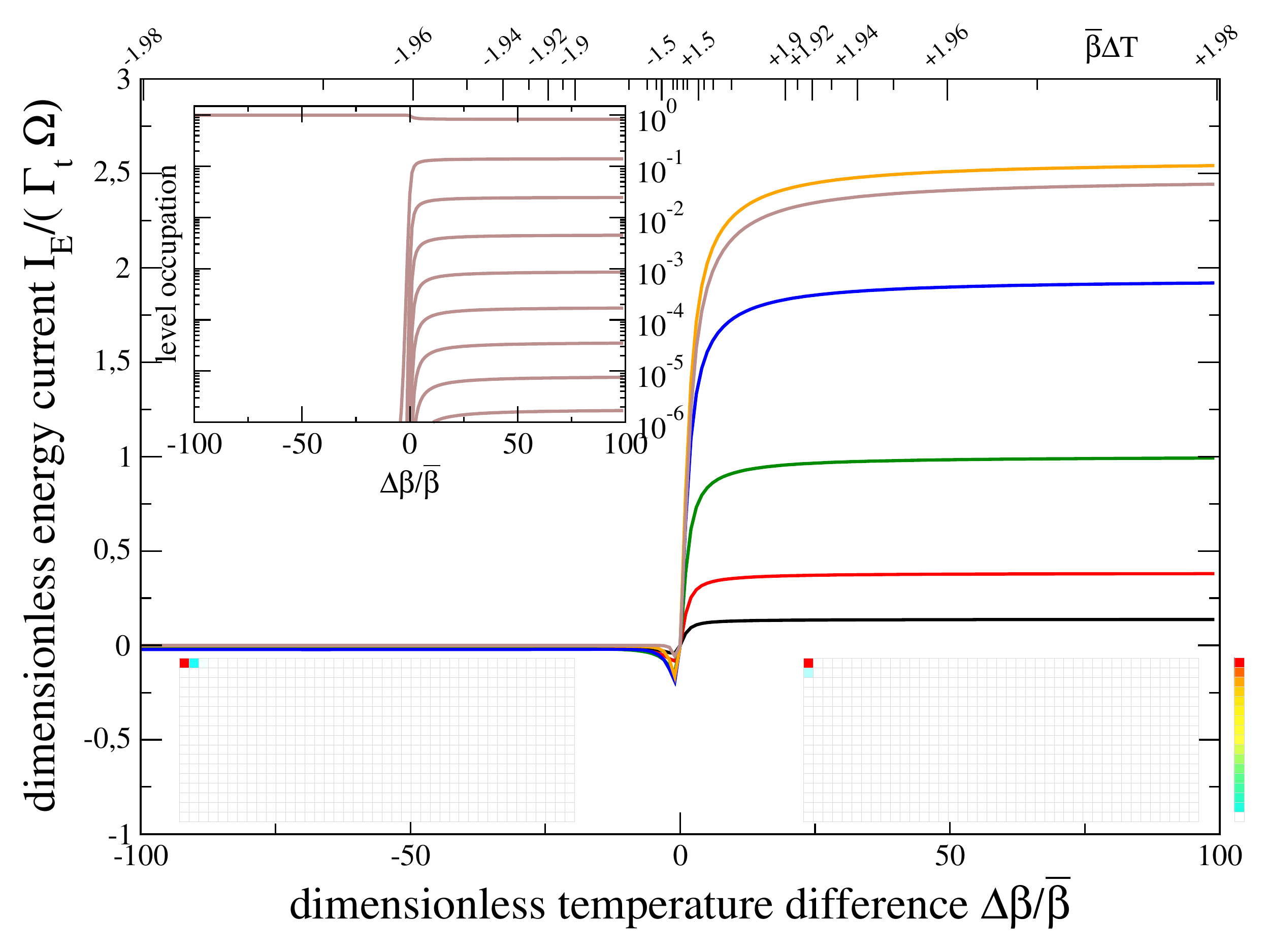}
\caption{\label{FIG:energycurrentb}(Color Online)
Similar as Figs.~\ref{FIG:energycurrent} and~\ref{FIG:energycurrenta}, but for an even lower average 
temperature $\bar\beta\Omega=1.0$.
No quadratic scaling is observed, the current is significantly suppressed.
Furthermore, the current for $\Delta\beta/\bar\beta\gg 0$ does not even rise monotonically with $N$, 
but rectification is still present. 
Color codes and other parameters have been chosen as in Fig.~\ref{FIG:energycurrent}.
}
\end{figure}

When we further decrease the average temperature, compare Fig.~\ref{FIG:energycurrentb}, 
the currents are further reduced.
Furthermore, in the conducting direction it does not even rise monotonically with $N$.

\subsubsection{Strong-coupling Current}

An ideal heat diode should have a large current in the conducting direction and should faithfully block the current when the direction
is reversed.
It is therefore reasonable to probe the strong-coupling regime, as increasing $\lambda$ should naively also increase the current.
However, we note that for our model this is only partially true, as the increased level renormalization will also reduce the energy current.
In fact, previous investigations have found a suppression of transient superradiance in the strong-coupling-limit~\cite{vorrath2005a}.
We also find an analogous behaviour in the stationary regime.

\begin{figure}[t]
\includegraphics[width=0.48\textwidth,clip=true]{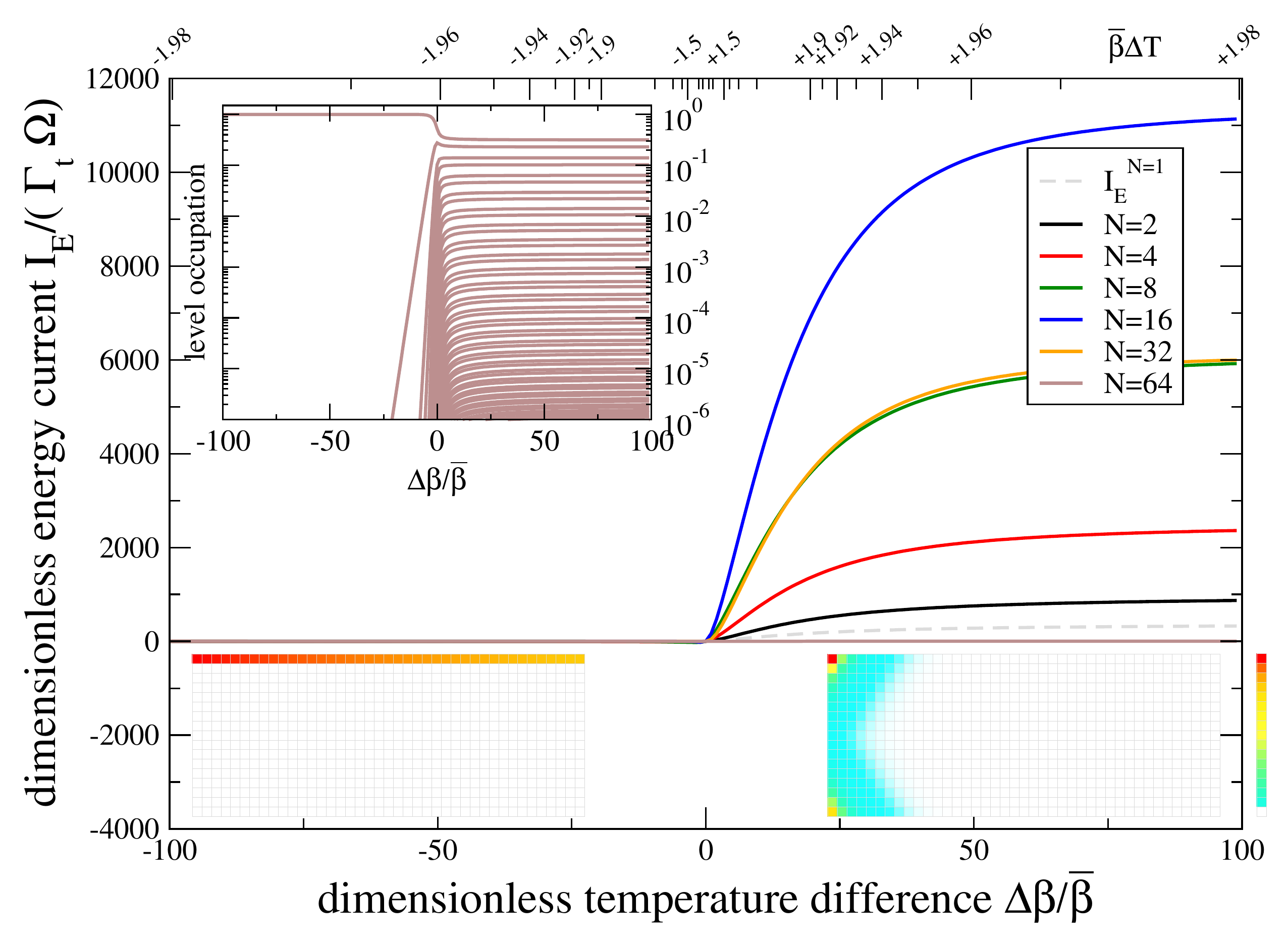}
\caption{\label{FIG:energycurrent_strong}(Color Online)
Plot of the dimensionless energy current between the reservoirs versus dimensionless temperature difference for 
strong coupling $\lambda=0.8\Omega$ and large average temperature $\bar\beta\Omega=0.01$.
The level renormalization prohibits for large $N$ the equipartition of levels (inset for $N=64$) for all non-equilibrium regimes 
and thus destroys the quadratic scaling of the current.
For small $N$ and $\Delta\beta\Omega \gg 0$ it grows approximately linearly with $N$ but for larger $N$ it is 
even reduced and above $N=64$ suppressed completely.
Nevertheless, for finite $N$ the quality of heat rectification is improved in comparison to the weak-coupling limit as
for negative $\Delta \beta$ all but the ground
state are exponentially suppressed, which directly affects the current.
Color coding and other parameters chosen as in Fig.~\ref{FIG:energycurrent}.
}
\end{figure}

For stronger couplings, the heat-diode capability is in principle even enhanced and also present for smaller temperature differences, 
since the stationary state becomes rapidly pure for $\Delta\beta \Omega < 0$ and thus effectively inhibits transport, 
see the inset of Fig.~\ref{FIG:energycurrent_strong}.
However, we also observe that for $\Delta\beta \Omega > 0$ the quadratic scaling of the current does not hold over the complete range of $N$.
In fact, the current is for large $N$ (in the Figure for $N=32$ and $N=64$) even further suppressed, which limits the throughput capability of 
the heat diode.
This is a consequence of the level renormalization, 
which destroys the previously observed equipartition of all energy levels (see inset).
We note that in the strong-coupling limit, the current in conductance direction is carried by two non-communicating regions in phase-space
(compare right density plot).
The mesostates with $m\approx 0$ are hardly occupied and do not contribute to the current. 
Instead, the system is rather concentrated close to the ground state.
Nevertheless, due to the strong coupling a significant current is produced for finite $N$.

When we lower the average temperature (increase $\bar\beta$) as in the weak-coupling regime, the total current is strongly suppressed without 
substantial changes in heat rectification properties (not shown).
These regimes are therefore less useful for heat diode application purposes.

We note that the diode effect requires $\Gamma_\ell \gg \Gamma_t$ 
(as one may have expected from the violation of the detailed balance relation due
to coarse-graining) and that one of the temperatures is small in comparison to the system energy scales
(to concentrate the populations at the boundaries of the phase space).
%

\section{Generality of the heat-diode effect}\label{SEC:rectification}

The pure-dephasing character of the interaction between the large spin and the collective mode facilitates the analytic diagonalization
of our system but also leaves open the question of what are the fundamental prerequisites to produce the heat diode behavior.
In the previous sections, we have identified a configuration blockade mechanism as being responsible for the blockade of the heat
current in one direction.
This picture is well confirmed in the density plots in Figs.~\ref{FIG:energycurrent},~\ref{FIG:energycurrenta},~\ref{FIG:energycurrentb}, 
and~\ref{FIG:energycurrent_strong}.
However, a configuration space similar to Fig.~\ref{FIG:rategraph_sketch} should arise in any bipartite system with 
couplings that connect to the individual constituents locally (on the level of the Hamiltonian).
Whenever the coupling to one reservoir is much stronger than that to the other reservoir and also than the coupling between the constituents, 
we will obtain a similar scenario: The heat current will be suppressed when the strongly-coupled reservoir is hot
and the weakly-coupled reservoir is cold.
The super-transmittant amplification of the current in the throughput direction affects the quantitative 
performance of the heat diode.

\begin{figure}[ht]
\includegraphics[width=0.48\textwidth,clip=true]{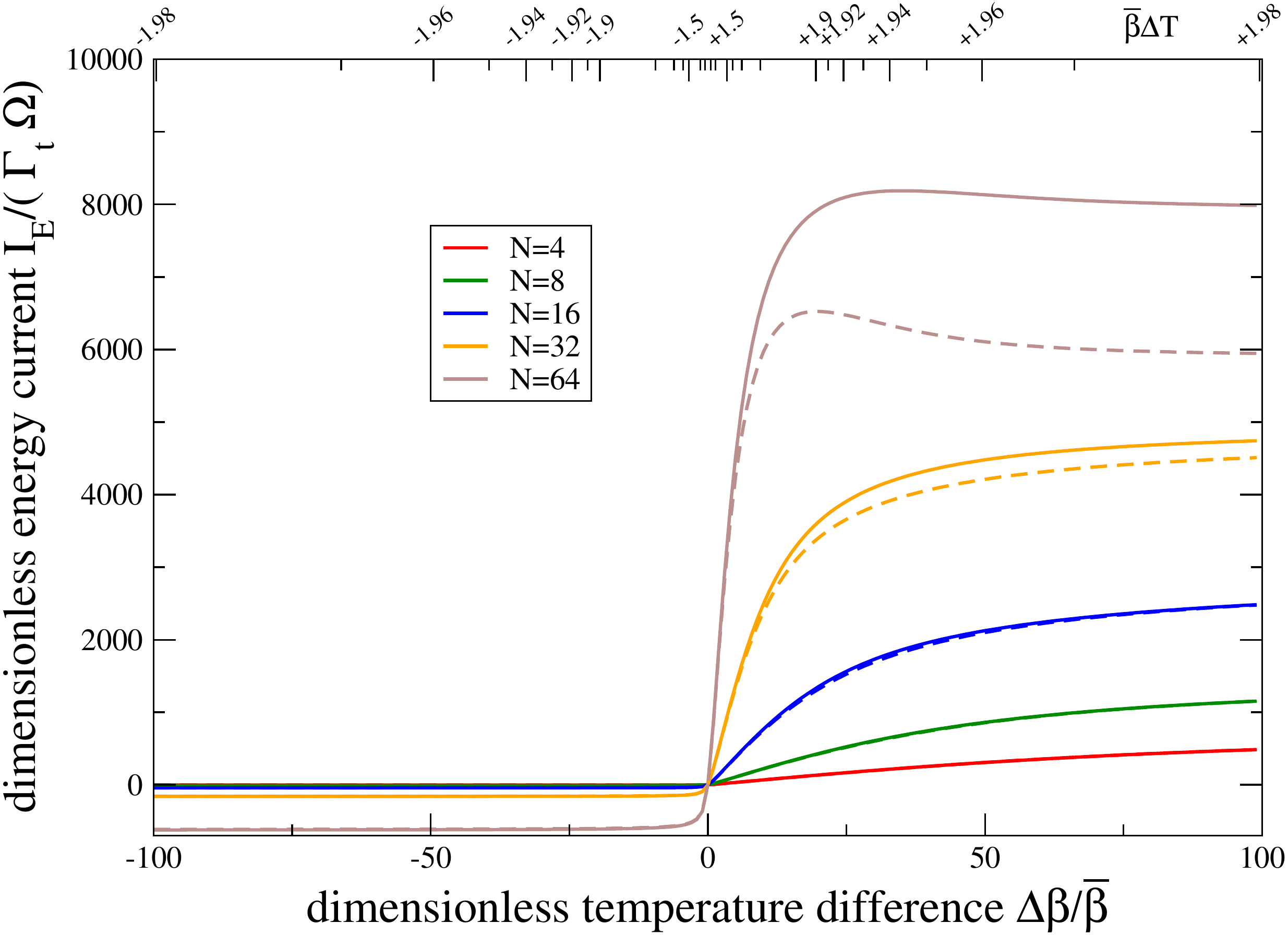}
\caption{\label{FIG:energycurrentjx}(Color Online)
Plot of the dimensionless energy current between the reservoirs versus dimensionless temperature difference for a modified model, 
where in Eq.~(\ref{EQ:ham_sys}) $\lambda J_z \to \lambda J_x$, without any coarse-graining approximation.
Solid curves correspond to a bosonic cutoff $N_{\rm cut}=10$ and dashed curves of similar color to a bosonic cutoff $N_{\rm cut}=20$,
showing that for small $N$ the cutoff has little effect in this regime.
Color coding and other parameters chosen as in Fig.~\ref{FIG:energycurrent}.
}
\end{figure}

To confirm this hypothesis, we have replaced the internal coupling in Eq.~(\ref{EQ:ham_sys}) by a dissipative one, $\lambda J_z \to \lambda J_x$.
Then, the system Hamiltonian implements the closed Dicke model~\cite{dicke1954a}, which is a well-known toy model for a quantum-critical system~\cite{tavis1968a,hepp1973a,wang1973a}.
The model can be mapped to coupled harmonic oscillators by employing a Holstein-Primakoff transformation, amenable to further simplifications in the large $N$-limit.
However, to compare with our previous calculations we are interested in the finite-$N$ limit, where a numerical approach is advisable (the spectral corrections due to 
first order perturbation theory in $\lambda$ have no effect on the rate equation).
The new Hamiltonian defines a new energy eigenbasis (computed numerically for a finite bosonic cutoff $N_{\rm cut}$), within which we derive
a rate equation similar to Eq.~(\ref{EQ:full_rate_equation}) and Eq.~(\ref{EQ:full_transition_rates}).
Here, the only difference is that the eigenstates and eigenvalues have to be obtained numerically and are therefore characterized by a single index and
not as before by angular momentum $m$ and boson occupation $n$.
This also implies that a coarse-graining procedure is not straightforward, such that the full model has to be solved numerically (as was done for the symbols in Fig.~\ref{FIG:energycurrent}).
The results depicted in Fig.~\ref{FIG:energycurrentjx} show that the rectification effect is still present and very pronounced, confirming our conjecture.

\section{Conclusions and Perspectives}\label{SEC:summary}

We have studied non-equilibrium physics in an ensemble of $N$
identical two-level systems asymmetrically coupled with two different
reservoirs. 
One is a standard heat bath, while the other reservoir is
structured, with a single bosonic mode coupled collectively to the
the ensemble of $N$ two-level systems and also to it its own heat bath.
By taking the evolution of the single bosonic mode dynamically into account, we can
model the strong-coupling limit with that reservoir.
We mainly considered the case where the coupling with the standard reservoir is dissipative
(transversal coupling) while the coupling with the structured reservoir is
assumed to be purely dephasing (longitudinal coupling). 
We showed that our results also apply to the case in which
both couplings are locally dissipative, but one reservoir is still structured.
For large coherence lengths in the reservoirs, the
coupling is collective and the ensemble of two-level systems behaves as
a single large spin. 

From the technical point of view, we mention that in the limit where the coupling between the longitudinal boson mode and its
reservoir is much larger than the coupling of the large spin to the transversal reservoir, we can approximately coarse-grain the dynamics by considering 
a reduced master equation for the large-spin eigenstates only.
This comes with the advantage of a tremendously reduced numerical complexity while maintaining the possibility of a thermodynamic
interpretation.
In Appendix~\ref{APP:alternative_derivation} we demonstrate that the coarse-grained description 
generally applies when the longitudinal mode is forced to remain in a thermal state.

Our main results can be thus summarized as follows:

{\it Superradiance.}
In usual investigations of superradiance only the coupling with the transverse reservoir is considered. 
There, the system starts from the largest angular momentum state (corresponding to all two-level systems inverted). 
While the system relaxes to the ground state, it passes through superradiant states which emit radiation with an intensity 
proportional to the number of atoms squared $N^2$. 
We have investigated the fate of superradiance in an equilibrium environment, where both the longitudinal and transverse reservoirs were held at zero temperature. 
For strong longitudinal coupling strength and/or large system sizes $N$, superradiance can be strongly affected by the presence of the longitudinal bath.
While the presence of an additional decay channel may in some parameter regimes enhance the relaxation speed, the longitudinal reservoir also
induces obstacles to superradiance, essentially due to modifications in the system energy levels: 
First, the conventional largest angular momentum state is for large system sizes and/or strong couplings no longer the energetically most excited state.
This implies that to relax toward the true ground state, the system would have to tunnel through a huge energy barrier, leading to an exponential suppression of
relaxation and thereby a complete superradiance blockade in this regime. 
We have circumvented this problem by choosing an initial state that in principle enables a fast relaxation toward the true ground state.
Second, one observes that the energy distance between the low energy states can become very large for large couplings and/or large system sizes.
Depending on the details of the reservoirs (technically expressed by their spectral coupling densities), they may not be able to absorb 
such large energies, which may also strongly suppress the final stages of superradiance.

{\it Rectification and Supertransmittance.}
We also investigated the non-equilibrium dynamics by keeping the reservoirs at different temperatures.
This gives rise to a stationary heat current through the system.
Now, depending on the stationary state the system assumes, the current can display very different 
scalings with the system size $N$.
While the current from a hot transverse reservoir to a cold longitudinal reservoir is supertransmittant (i.e., scales as $N^2$ in the
weak-coupling regime), for the opposite thermal bias the current is strongly reduced and becomes almost independent of $N$.
This effectively implements a heat diode, with a rectification factor
that can be tuned by changing the system size $N$.
The essential features needed to obtain the heat
diode effect in a generic system are discussed in Sec.~\ref{SEC:rectification}.

Our setup constitutes a proof of principle of how extremely large rectification factors  can be achieved by exploiting collective
couplings with thermal reservoirs. 
We have argued that our model could be used to describe trapped ions collectively 
coupled to a thermal photon field and a phonon field. 
Since rectification allows energy transfer from the hot photon field (transverse) to the cold phonon field (longitudinal), 
these findings could inspire the design of a device able to efficiently absorb energy from sunlight 
(that corresponds to a black-body radiator with high temperature of $\approx 6000$ K) and convert it efficiently (due to supertransmittance) 
into heat stored in a phonon reservoir.
In the absence of sunlight, the inverse process would be strongly suppressed, such that the total device would be
a very suitable energy harvester.
We expect our findings to apply to transport through bipartite systems with highly asymmetric couplings, 
and the design of such devices is an appealing avenue of further research.


\section{Acknowledgments}

Financial support by the DFG (grant SCHA \mbox{1646/3-1}) and helpful discussions with T. Brandes and P. Strasberg are gratefully acknowledged.
G. G. G. gratefully acknowledges the support of the Mathematical Soft Matter Unit of the Okinawa Institute of Science and Technology Graduate University. 

\bibliographystyle{unsrt}
\bibliography{references}

\appendix


\section{Stationary State}\label{APP:stationary_state}

To calculate the heat current, we first need to calculate the steady state of the rate equation.
Numerically, we have found that the determination of the null space is not always stable.
Therefore, we determined the null space of a rate matrix by computing the adjugate matrix
via the transpose of the cofactor matrix.
In case of a tri-diagonal rate matrix
\bea
({\cal M})_{ij} &=& \delta_{j,i+1} m_{i,i+1} + \delta_{j,i-1} m_{i,i-1}\nn
&& -\delta_{ij}(m_{i-1,i}+m_{i+1,i})
\eea
with dimension $(N+1)\times (N+1)$ this calculation simplifies considerably.
One can then check that the steady state is given by (we use $M=N+1$ for generality)
\bea
\bar{P}_k = \frac{\left[\prod\limits_{i=2}^{k} m_{i,i-1}\right] \left[\prod\limits_{j=k}^{M-1} m_{j,j+1}\right]}
{\sum\limits_k \left[\prod\limits_{i=2}^{k} m_{i,i-1}\right] \left[\prod\limits_{j=k}^{M-1} m_{j,j+1}\right]}\,,
\eea
where $k\in\{1,\ldots,M\}$.


\section{Alternative derivation of the coarse-grained rate equation}\label{APP:alternative_derivation}

Here, we will show that we can also obtain the coarse-grained rate equation~(\ref{EQ:modified_rate_equation}) from a model where
the longitudinal boson is not coupled to an independent reservoir, i.e., where the total Hamiltonian simply reads
\bea
H &=& \frac{\omega_0}{2} J_z + J_z (\lambda a + \lambda^* a^\dagger) + \Omega a^\dagger a\nn
&&+ J_x \sum_k \left(h_k b_k + h_k^* b_k^\dagger\right) + \sum_k \omega_k b_k^\dagger b_k\,.
\eea
To treat the model within a master equation approach, we consider only the large spin as the system, 
and to treat the strong-coupling limit, too, we use the polaron transformation~(\ref{EQ:polaron_transformation}).
With Eq.~(\ref{EQ:polaron_action}), we conclude that 
under a polaron transformation, the Hamiltonian transforms according to
\bea\label{EQ:ham_polaron_transformed}
U H U^\dagger &=& \frac{\omega_0}{2} J_z - \frac{\abs{\lambda}^2}{\Omega} J_z^2 + \Omega a^\dagger a + \sum_k \omega_k b_k^\dagger b_k\\
&&+ \left(J_+ e^{+2 B} + J_- e^{-2 B}\right) \sum_k \left(h_k b_k + h_k^* b_k^\dagger\right)\,.\nonumber
\eea
Thus, the coupling between spin and longitudinal mode goes away at the expense of a dressed spin-boson coupling.

\begin{figure}
\includegraphics[width=0.48\textwidth,clip=true]{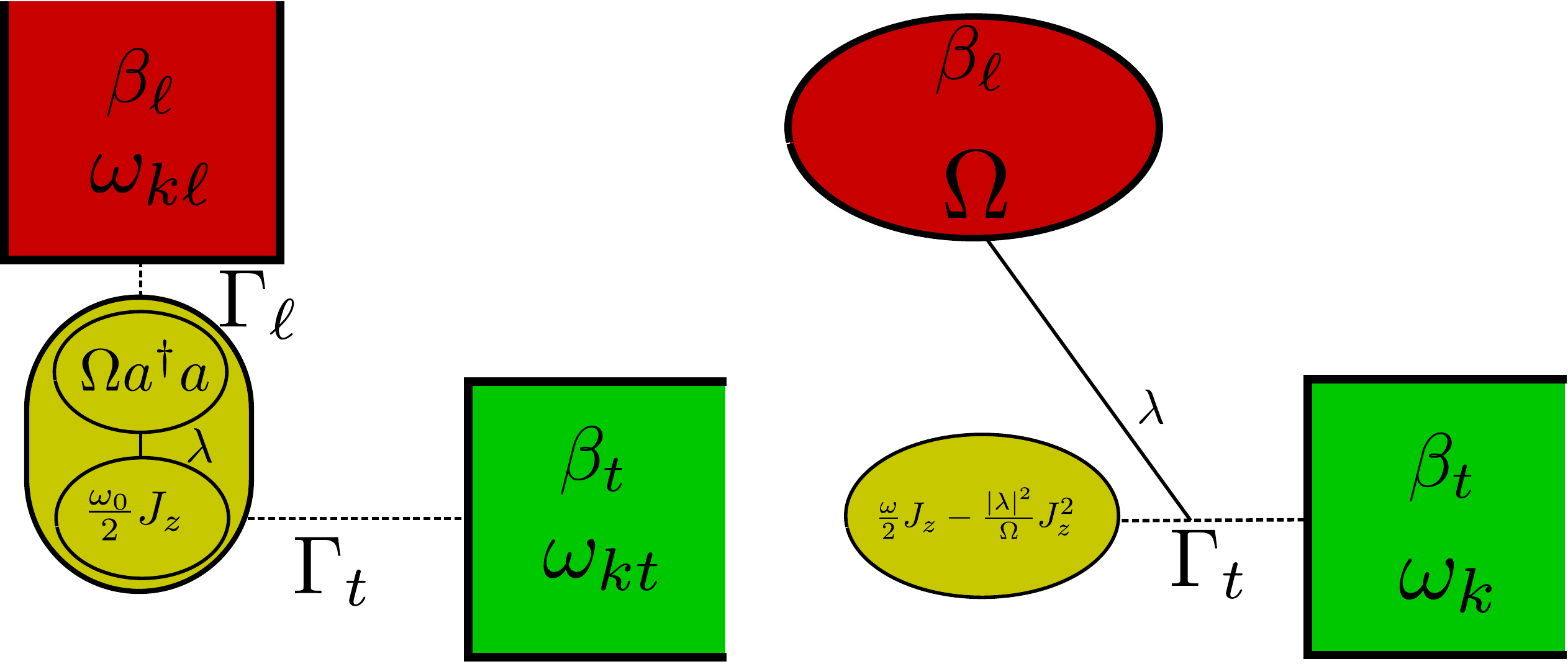}
\caption{\label{FIG:sketch_polaron}(Color Online)
Putting the longitudinal boson mode in a thermal state in the polaron frame (right, appendix) leads to the same evolution equation as 
coupling it to a separate continuous bosonic reservoir (left, main manuscript) and then assuming that the longitudinal boson degrees of 
freedom relax to their equilibrium state much faster than the large spin, such that a reduced coarse-grained  Markovian description only in 
terms of the large spin eigenstates applies.
}
\end{figure}

We note that we can derive a master equation with standard methods that is perturbative in $h_k$ but non-perturbative in $\lambda$.
In doing so, we will put both the longitudinal boson mode and the bosons in thermal equilibrium states with inverse temperatures
$\beta_\ell$ and $\beta_t$, respectively~\cite{schaller2013a,krause2015a}.
Since the polaron transformation is non-local between large spin and longitudinal mode, simply placing the boson mode
in a thermal state does not correspond to a simple thermal state in the original frame.
Instead, its state becomes conditioned on the large spin state~\cite{schaller2014}.
Here, we will show that the resulting rate equation is identical to the one obtained
in the main paper via coarse-graining~(\ref{EQ:modified_rate_equation}), see also Fig.~\ref{FIG:sketch_polaron}.

Evidently, the eigenenergies of the system Hamiltonian in Eq.~(\ref{EQ:ham_polaron_transformed}) are given by~(\ref{EQ:spectrum_spin}), and 
with identifying the coupling operators as $A_1 = J_+$, $B_1 = e^{+2 B} \sum_k (h_k b_k + h_k^* b_k^\dagger)$, $A_2 = J_-$, and
$B_2 = e^{-2 B} \sum_k (h_k b_k + h_k^* b_k^\dagger)$, we can set up a rate equation for the evolution of populations in the 
spin energy eigenstates.
To do so, we have to evaluate the matrix elements of the system coupling operators -- using Eqns.~(\ref{EQ:ladder_operators}) -- which imply that only two reservoir
correlation functions have to be found to evaluate the rate from energy eigenstate $b$ to energy eigenstate $a$
\bea\label{EQ:rate_reduced}
\gamma_{ab,ab} &=& \gamma_{12}(E_b-E_a) \abs{\bra{a} J_- \ket{b}}^2\nn
&& + \gamma_{21}(E_b-E_a) \abs{\bra{a} J_+ \ket{b}}^2\,.
\eea

Consequently, we calculate the reservoir correlation function for the reservoir coupling operators
\bea
{\cal B}^\pm = e^{\pm 2 B} \sum_k \left(h_k b_k + h_k^* b_k^\dagger\right)\,,
\eea
which enter the correlation functions in the form (bold symbols indicate the interaction picture)
\bea
\expval{\f{\cal B}^\pm(\tau) {\cal B}^\mp} &=& C^{\pm}_\ell(\tau) C_t(\tau)\,,\nn
C^{\pm}_\ell(\tau) &=& \expval{e^{\pm 2 \f{B}(\tau)} e^{\mp 2 B}}\,,\nn
C_t(\tau) &=& \frac{1}{2\pi} \int_{-\infty}^0 \Gamma_t(-\omega) n_t(-\omega) e^{-\ii\omega\tau} d\omega\\
&& + \frac{1}{2\pi} \int_0^{+\infty} \Gamma_t(+\omega) [1+n_t(+\omega)] e^{-\ii \omega \tau} d\omega\,.\nonumber
\eea
At this state it is already evident that the resulting rates will not be additive in the two reservoirs.

The longitudinal contributions can be written as
\bea
C_\ell^{+}(\tau) &=& e^{-\frac{4\abs{\lambda}^2}{\Omega^2}\left[\left(1-\cos(\Omega \tau)\right)\coth\left(\frac{\beta_{\ell}\Omega}{2}\right) + \ii \sin(\Omega \tau)\right]}\,,\nn
C_\ell^{-}(\tau) &=& C_\ell^{+}(\tau) \equiv C_\ell(\tau)\,,
\eea
and it is visible that these do not decay to zero at infinity.
One might be tempted to consider this as problematic with regard to the Markovian approximation.
However, the longitudinal correlation function always enters in product form with the transversal 
correlation functions, such that the total correlation function always decays.
To interpret their action in a more physical way we rewrite the correlation functions as
\bea
C_\ell(\tau) &=& e^{-\frac{4\abs{\lambda}^2}{\Omega^2}\left[ (1+2n_\ell) - n_\ell e^{+\ii \Omega\tau} - (1+n_\ell) e^{-\ii\Omega\tau}\right] }\,,
\eea
where $n_\ell = [e^{\beta_\ell \Omega}-1]^{-1}$.
We can easily check their KMS relations $C_\ell(\tau) = C_\ell(-\tau-\ii\beta_\ell)$.
We can compute the Fourier transform of the longitudinal mode correlation function by formally expanding in powers of $e^{\pm \ii \Omega \tau}$
\bea\label{EQ:polaron_correlation_function}
\gamma_\ell(\omega) &=& \int C_\ell(\tau) e^{+\ii\omega\tau} d\tau\nn
&=& 2\pi e^{-\frac{4\abs{\lambda}^2}{\Omega^2} (1+2n_\ell)}\sum_{m,m'=0}^\infty \left[\frac{4\abs{\lambda}^2}{\Omega^2}\right]^{m+m'}\times\nn
&&\times \frac{n_\ell^m [1+n_\ell]^{m'}}{m! m'!} \delta(\omega-(m'-m)\Omega)\\
&=& 2\pi e^{-\frac{4\abs{\lambda}^2}{\Omega^2} (1+2n_\ell)} \sum_{\bar{n}=-\infty}^{+\infty} \delta(\omega - \bar{n} \Omega) \times\nn
&&\times \left(\frac{1+n_\ell}{n_\ell}\right)^{\bar{n}/2} {\cal J}_{\bar{n}}\left(\frac{8\abs{\lambda}^2}{\Omega^2} \sqrt{n_\ell(1+n_\ell)}\right)\,,\nonumber
\eea
where ${\cal J}_n(x)$ denotes the modified Bessel function of the first kind.

The bosonic contribution has standard form and also obeys a KMS condition of the form $C_t(\tau) = C_t(-\tau-\ii\beta_t)$.

The full Fourier transform of the correlation function is given by 
\bea
\gamma(+\omega) &=& \int C_\ell(\tau) C_t(\tau) e^{+\ii\omega\tau} d\tau\,,
\eea
and we note that we can represent these also by convolution integrals of the separate Fourier transforms
\bea
\gamma(\omega) = \frac{1}{2\pi} \int \gamma_\ell(\omega-\bar\omega) \gamma_t(\bar\omega) d\bar\omega\,.
\eea
Inserting Eq.~(\ref{EQ:polaron_correlation_function}) eventually yields 
\bea\label{EQ:total_correlation_function}
\gamma(\omega) &=& \sum_{\bar{n}=-\infty}^{+\infty} \gamma_{\bar{n}}(\omega)\,,\\
\gamma_{\bar{n}}(\omega) &=& \alpha_{\bar{n}} \gamma_{11}(\omega-\bar{n} \Omega)\,,\nonumber
\eea
where $\gamma_{11}(\omega)$ is defined in Eq.~(\ref{EQ:ft_corrfunc}) in the main manuscript.

Inserting these results in Eq.~(\ref{EQ:rate_reduced}), we find that the resulting rate equation is identical with Eq.~(\ref{EQ:modified_rate_equation}) in the main manuscript.

Independent calculations have shown that coarse-graining approaches also exist for previously treated electron-phonon models~\cite{schaller2013a,krause2015a,strasberg2016a} (not shown).

\end{document}